\documentclass[fleqn,usenatbib]{mnras}
\usepackage[T1]{fontenc}
\usepackage{ae,aecompl}
\usepackage{graphicx}
\usepackage{amsmath}
\usepackage{amssymb}
\usepackage{mathrsfs}
\usepackage{lmodern}

\usepackage{xcolor}

\usepackage{hyperref}
\usepackage[hyphenbreaks]{breakurl}
\usepackage{url}

\renewcommand{\d}[1]{\ensuremath{\operatorname{d}\!{#1}}}

\newcommand{\degC}[1]{#1\,^\circ\rm{C}}

\newcommand\Tcmb{T_{\rm{CMB}}}

\newcommand\Tatm{T_{\rm{atm}}}

\newcommand\tatm{t_{\rm{atm}}}
\newcommand\Iatm{I_{\nu,\rm{atm}}}

\newcommand\TSL{T_{\rm{SL}}}

\newcommand\pH{p_{\rm{H}}}
\newcommand\xv{x_{\rm{v}}({\rm H_2O})}
\newcommand\trpwv{{\em TR-PWV-DR-1}}

\newcommand{\Beginruledtabular}{\begin{center}\begin{small}}
\newcommand{\Endruledtabular}{\end{small}\end{center}}

\newcommand{\BeginruledtabularSec}{\hline}
\newcommand{\EndruledtabularSec}{\hline}
\newcommand{\footnotetextTAB}[1]{{\small #1}\\}
\newcommand{\footnotemarkTAB}[1]{{#1}}

\providecommand{\adsurl}[1]{\href{#1}{ADS}}

\title[Clear sky atmosphere from climatology data]{Clear sky atmosphere at cm-wavelengths from climatology data}

\author[B. Lew, et al.]{
Bartosz Lew,$^{1}$\thanks{E-mail: blew@astro.uni.torun.pl}
Joanna Uscka-Kowalkowska,$^{2}$
\\
$^{1}$Toru\'n Centre for Astronomy, Nicolaus Copernicus University, ul. Gagarina 11, 87-100 Toru\'n, Poland\\
$^{2}$Department of Meteorology and Climatology, Nicolaus Copernicus University, Lwowska 1, 87-100 Toru\'n, Poland
}

\date{Accepted 2015 October 28.  Received 2015 October 08; in original form 2015 July 02}

\pubyear{2015}

\begin{document}
\label{firstpage}
\pagerange{\pageref{firstpage}--\pageref{lastpage}}
\maketitle

\begin{abstract}
We utilise ground-based, balloon-borne and satellite climatology data
to reconstruct site and season-dependent vertical profiles of
precipitable water vapour (PWV).  We use these profiles to 
solve radiative transfer through the atmosphere, and derive 
atmospheric brightness temperature ($\Tatm$) and
optical depth ($\tau$) at centimetre wavelengths.

We validate the reconstruction by comparing the model column PWV with
photometric measurements of PWV, performed in clear sky conditions
pointed towards the Sun. Based on the measurements, we devise a selection
criteria to filter the climatology data to match the PWV levels to the
expectations of the clear sky conditions.

We apply the reconstruction to the location of a Polish 32-metre
radio telescope, and characterise $\Tatm$ and $\tau$ year-round, at
selected frequencies. We also derive the zenith distance dependence
for these parameters, and discuss the shortcomings of using planar,
single-layer, and optically thin atmospheric models in
continuum radio-source flux-density measurement calibrations.

We obtain PWV-$\Tatm$ and PWV-$\tau$ scaling relations in clear sky
conditions, and constrain limits to which the actual $\Tatm$ and $\tau$
can deviate from those derived solely from the climatological data.

Finally, we suggest a statistical method to detect clear sky that
involves ground-level measurements of relative humidity. Accuracy
is tested using local climatological data.  The method may be useful to
constrain cloud cover in cases when no other (and more robust)
climatological data are available.

\end{abstract}


\section{Introduction}
\label{sec:intro}
Ground-based, cm-wavelengths radio continuum observations of
astrophysical sources depend on atmospheric emission, and absorption
of the incident radiation.  Time varying line-of-sight (LOS) abundance
of ice, liquid water, and water vapour generates variable optical depth,
which leads to signal instabilities.  These instabilities do not
average out under long integrations due to the steep spectrum of
turbulent atmospheric water density fluctuations.  The atmospheric
effects associated with dry air thermal emission, continuum and line
absorption are frequency-dependent, and can be characterised by
atmospheric brightness temperature ($\Tatm(\nu)$) and transmittance
($\tatm(\nu)$).  These atmospheric emissions contribute to the system
temperature ($T_{\mathrm{sys}}$) that limits the sensitivity of any
ground-based telescope-receiver pair.

Over the last few decades, monitoring temperature, pressure, density,
and other altitude dependent parameters of air, have helped to develop
a few widely-accepted models of Earth's atmosphere. Advances in
atomic and molecular line spectroscopy provided absorption
coefficients for various gas species, and computer-generated spectra
for mixtures of gases under given thermodynamic conditions can now
reproduce these observations in great detail.  Thus, the radiative
properties of the atmosphere are well known, and can be derived from
the first principles, from radio waves to infrared frequencies.

At cm-wavelengths the electromagnetic spectrum of $\Tatm$ is
predominantly defined by continuum absorption, and emission of oxygen
($\rm O_2$) and water vapour (${\rm H_2O}$) molecules.  The latter
results from strong coupling of the electric dipole moment of water
molecules to the millimetre radiation via rotational
transitions. Going to liquid and ice states less and no rotational
freedom is possible respectively, and therefore droplets and ice
particles are expected to radiate less per molecule.  In the presence
of clouds, the background signals are additionally attenuated in ice
and droplets, decreasing the signal-to-noise ratio (SNR) of any radio
source. This translates to an increased observational time required to
detect the same source at the same significance level, as compared to
the situation without clouds.  Given that the cloud cover is
non-uniform and variable, the amount of the attenuation will change
over time, generating variance at different time scales in the signal
received through the main beam, or through side-lobes. For the radio
continuum flux density observations at cm- and mm-wavelengths, this
means that nearly clear sky conditions are required,
although in practice, observations are also viable when some
high-level icy clouds are present.

Unlike the '\emph{dry}' component of air, the distribution of
atmospheric water on Earth is strongly dependent on location. For this
reason many dedicated radio telescopes are built at high altitudes and
in dry and/or cold climates (such as the Atacama desert or the South
Pole). This reduces atmospheric emissions, maximises transmittance,
improves thermal stability, and rules out cloud attenuation. At other
locations, atmospheric water variations must be monitored to optimally
match the observational programme to the current weather conditions.
In clear sky conditions, measurements of precipitable water vapour
(PWV) can be useful in working out optical depths and estimating
atmospheric absorption corrections of astronomical radio source flux
density measurements.

Atmospheric water vapour can be measured in a number of ways, including
(i) radio sounding,
(ii) atmospheric delays of GPS-satellite signals, 
(iii) sun photometry of water lines in near-infrared light,
(iv) direct radiometric measurements in water bands (e.g. \cite{Liljegren2001}).
In the larger time scales, PWV can be modelled statistically using
climatological data.

Currently, the publicly-available data from ground-based
meteorological stations, radio sounding, and satellite observations
allow reconstruction of the vertical structure of atmosphere at
particular locations of interest.  For example, the Integrated Global
Radiosonde Archive (IGRA) data, which we use in this analysis,
provides substantial aerial coverage and density worldwide,
but only a few atmospheric parameters are
measured at relatively large time intervals (only twice a day).
Likewise, many satellite data obtained with the solar occultation method
also have nearly global coverage, but their sensitivity is largely
limited in the troposphere due to clouds. However, these meteorological data
can be used to model the total average PWV on a month-by-month
basis.  Whether by direct radiometric observations or by solving
radiative transfer equations, many radio astronomical observation
sites have $\Tatm$ and $\tatm$ calculated at the desired frequencies
and the time of year \citep{Ajello1995,Radford1998,Bussmann2005,
Bustos2014}.  In this paper we will characterise the atmosphere in
these terms for the first time for the location of the 32-metre radio
telescope (RT32) in Poland, and show that a similar approach can be
easily adapted to virtually any other location.  The RT32 is one of
the European VLBI Network stations, currently capable of observations
in the L, C, K and Ka frequency bands (see \cite{Lew2015} for more RT32
specifications).  

The motivation for investigating local atmosphere
arises from the planned radio source survey with the 30-GHz OCRA-f
focal plane, 4-pair, beam-switched receiver \citep{Lew2015}.  In
particular, we seek to improve the atmospheric model, previously used
for the continuum flux density measurement calibrations
\citep{Gawronski2010,Peel2010,Lancaster2011}, which requires estimates
of the atmospheric optical depth ($\tau$) at the source zenith
distance ($z_d$).  These estimates are typically obtained by performing tipping
scans, with the assumption that $\Tatm(z_d) \approx \Tatm(0)
\sec(z_d)$.  We investigate the limits of validity of this, and other
approximations, using radiative transfer in the atmosphere in clear
sky conditions.

Radio sounding and satellite data are recorded independently of
weather conditions, and modelling PWV in clear skies
requires employing selection criteria that would assure that the
filtered subset of the data corresponds to the clear sky conditions
at a particular location and observation time. A search for such
selection criteria, given the aforementioned sparsity and deficiencies
of the sounding data, is also one of the aims of this paper.

The structure of this paper is as follows.  In section~\ref{sec:data}
we discuss the publicly available climatological data used in our
reconstructions of atmospheric profiles, describe the atmospheric
model parametrisation and introduce our PWV measurements.  In
section~\ref{sec:transfer} we describe the radiative transfer approach
to calculating atmospheric transmittances and brightness
temperatures.  The main results are gathered in Sec.~\ref{sec:results}.
Discussion and conclusions are in sections~\ref{sec:discussion}
and~\ref{sec:conclusions} respectively.

\section{Data}
\label{sec:data}

We consider contributions to $\Tatm$ and $\tatm$ from dry air
(with standard nitrogen to oxygen proportions), ozone ($O_3$) and
water vapour (${\rm H_2O}$), and ignore atmospheric trace gases.
In the following subsections, we describe the climatological data,
used to reconstruct a localised, vertical structure of the atmosphere,
which is a starting point for radiative transfer.

\subsection{Vertical pressure and temperature profiles} 
\label{sec:CIRA}

\begin{figure}
\includegraphics[width=0.48\textwidth]{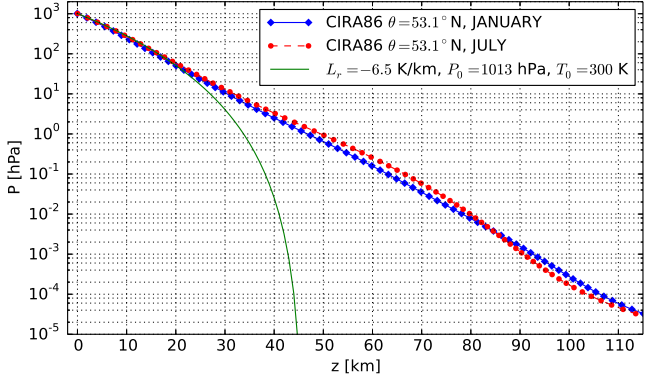}
\caption{Atmospheric pressure profiles in January (solid, blue) and in
July (dashed, red) interpolated from the CIRA86 data for the
latitude $\theta_T=53.1^\circ$N.  For comparison, the theoretical
profile (Eq.~\ref{eq:barometricFormulaCorrected})
is plotted for the assumed values of lapse-rate ($L_r$),
pressure ($P_0$), and temperature ($T_0$) (bottom
solid-green curve).}
\label{fig:cira86_pressure}
\end{figure}

\begin{figure}
\includegraphics[width=0.48\textwidth]{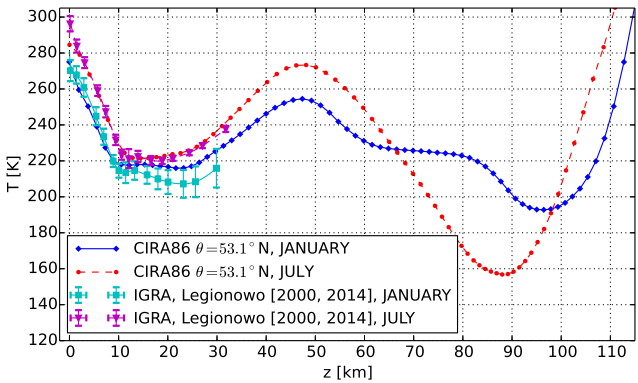}
\caption{Atmospheric temperature profiles in January (solid, blue) and
in July (dashed, red) interpolated from the CIRA86 data for the
latitude $\theta_T=53.1^\circ$N.  The curves extending up to
$z\approx 30$ km are the IGRA radiosonde data from Legionowo station,
averaged over the period of 15 years.  The horizontal error bars
represent $\pm 1\sigma$ dispersion resulting from
variable geopotential altitudes assigned to the fixed pressure
levels. The vertical error bars represent $\pm 1\sigma$
dispersion of temperatures in the selected month.  }
\label{fig:cira86_temperature}
\end{figure} 

We reconstruct vertical atmospheric temperature and pressure
profiles using the COSPAR\footnote{Committee on Space Research }
International Reference Atmosphere project
data\footnote{\protect\url{http://badc.nerc.ac.uk/data/cira}}
(hereafter CIRA86) \citep{CIRA86_0,CIRA86_1,CIRA86_2}.  These
zonally-averaged data are a compilation
of ground-based, radiosonde and satellite measurements of atmospheric
pressure, temperature, wind velocity and geopotential height up to an altitude
of $z=120$ km, with $5^\circ$ latitude resolution, roughly $2$ km
vertical resolution, and nearly global coverage in latitudes
($\theta=[80\,^\circ\mathrm{S},80\,^\circ\mathrm{N}]$). We use the
monthly averaged vertical profiles of pressure $P(z_g)$ and temperature
$T(z_g)$, given as a function of geopotential height and we apply
an interpolation to obtain a profile for the desired latitude.
Throughout this paper we will interpolate the
profiles for the latitude $\theta_T=53.1^\circ$N -- the latitude of
the Toru\'n Centre for Astronomy (TCfA).
Geopotential height ($z_g$) is
related to geometrical height ($z$) above the sea level via:
$  z_g = z\, R_\oplus / (R_\oplus+z) $,
where $R_\oplus=6371$ km is the assumed value of Earth radius.  In
what follows, we will ignore the difference between the two heights, as
they are almost identical over the range of altitudes of
interest. The differences are also unimportant when compared the to
variance generated by weather related pressure variations.
Associating the fixed CIRA86 pressure levels to altitudes ($z$) allows
us to assign geometrical heights to other physical quantities of
interest as well, such as temperature, relative humidity, and mixing ratios,
which are typically recorded at fixed pressure levels.

The interpolated CIRA86 pressure profiles are
shown in Fig.~\ref{fig:cira86_pressure}.
For visualisation we compare them against the theoretical model
with the altitude-independent lapse rate \citep{USstandardAtm1976}:
\begin{equation}
P(z) = P_0 \bigl[\frac{T_0}{T_0+ L_r(z-z_0)}\bigr]^{\frac{g_0 \mu}{R L_r}}
\label{eq:barometricFormulaCorrected}
\end{equation} where $L_r$ is the lapse rate (Fig.~\ref{fig:cira86_pressure}),
where $g_0=9.80665\, \mathrm{km/s^2}$ is the assumed surface
acceleration of the Earth,
$R=8.31432\, \rm{J\cdot  mol^{-1}\cdot K^{-1}}$ is the gas constant,
$\mu=0.0289644\, \rm{kg\cdot mol^{-1}}$
is the molar mass of the air and $P_0$, $T_0$ and $z_0$
are fiducial pressure, temperature and altitude respectively.
While the consistency is very good
in the troposphere, the deviation from the measured profiles in the
stratosphere results from
the lapse rate increasing to zero as one approaches tropopause,
the fact that we do not take into account in this visualisation.

In order to adjust the pressure profiles to the local conditions and
improve the reconstruction at low altitudes ($z\lesssim 13$ km), we
combine CIRA86 profile with the profile extracted from the local
sounding data (see section~\ref{sec:IGRA}).  However, the two profiles
turn out to be quite compatible, with the biggest discrepancy of $\sim
8\%$ at $z=30$ km in January, and much smaller in July.

Similarly, in the case of
temperature profiles, we gauge the impact of the longitudinal deviations
from the CIRA86 mean, by comparing it with the daytime
radio sounding profile from Legionowo (near Warsaw, Poland),
averaged over the period 2000-2014.  We find the two datasets
consistent to within 8\% in the stratosphere and in the lower
troposphere (Fig.~\ref{fig:cira86_temperature}).
The consistency is better between the two regimes. The
near-ground departures from the zonal averages are somewhat expected,
which is why we rely on the sounding data at low altitudes.  The two
datasets clearly mark the transition to tropopause at an altitude of
about 10 km at the considered latitude.

\subsection{Vertical WV profiles}
\label{sec:WVprofiles}

\subsubsection{Relative Humidity profiles}
\label{sec:IGRA}
\begin{figure*}
\centering
\includegraphics[width=0.48\textwidth]{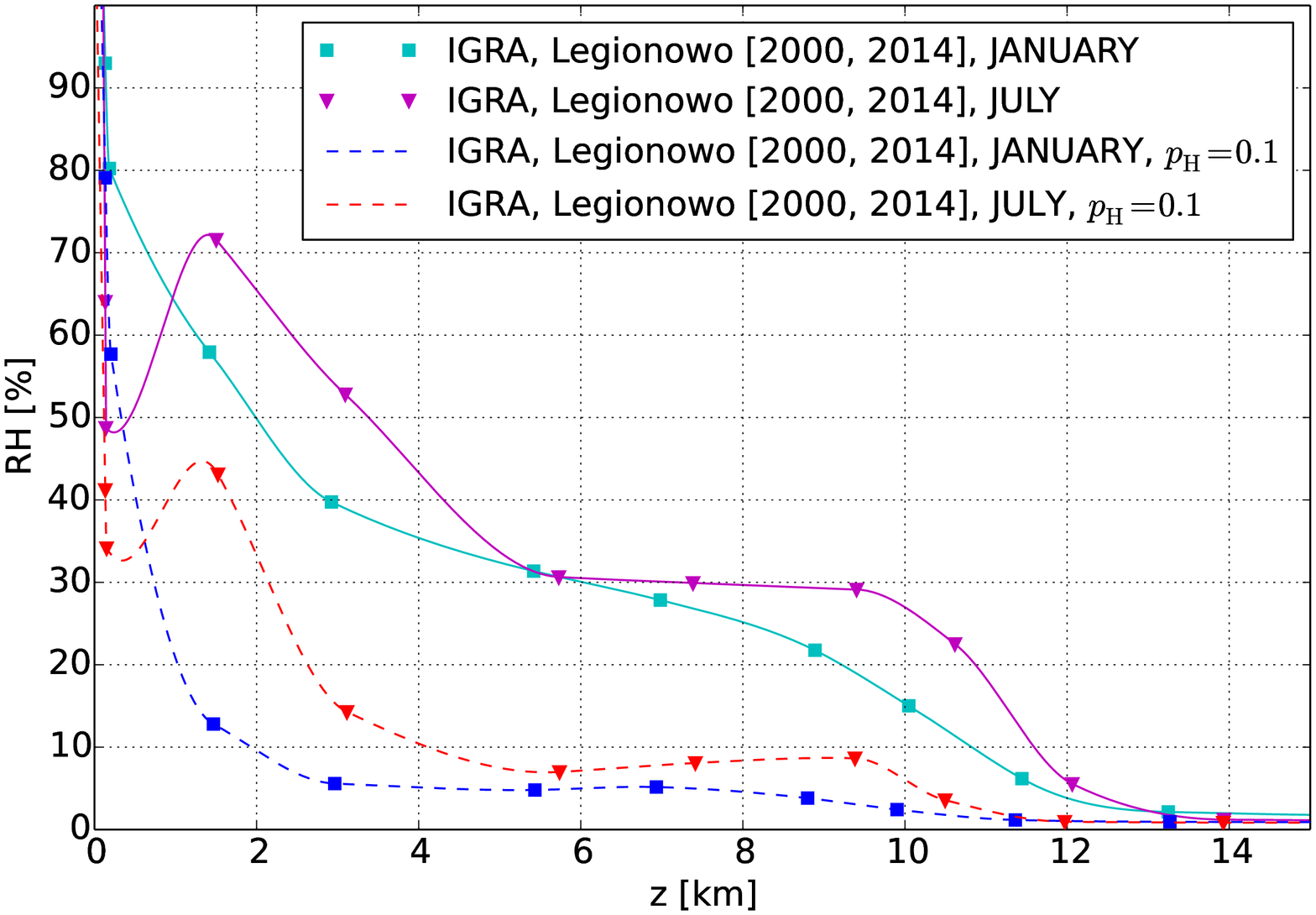}
\includegraphics[width=0.48\textwidth]{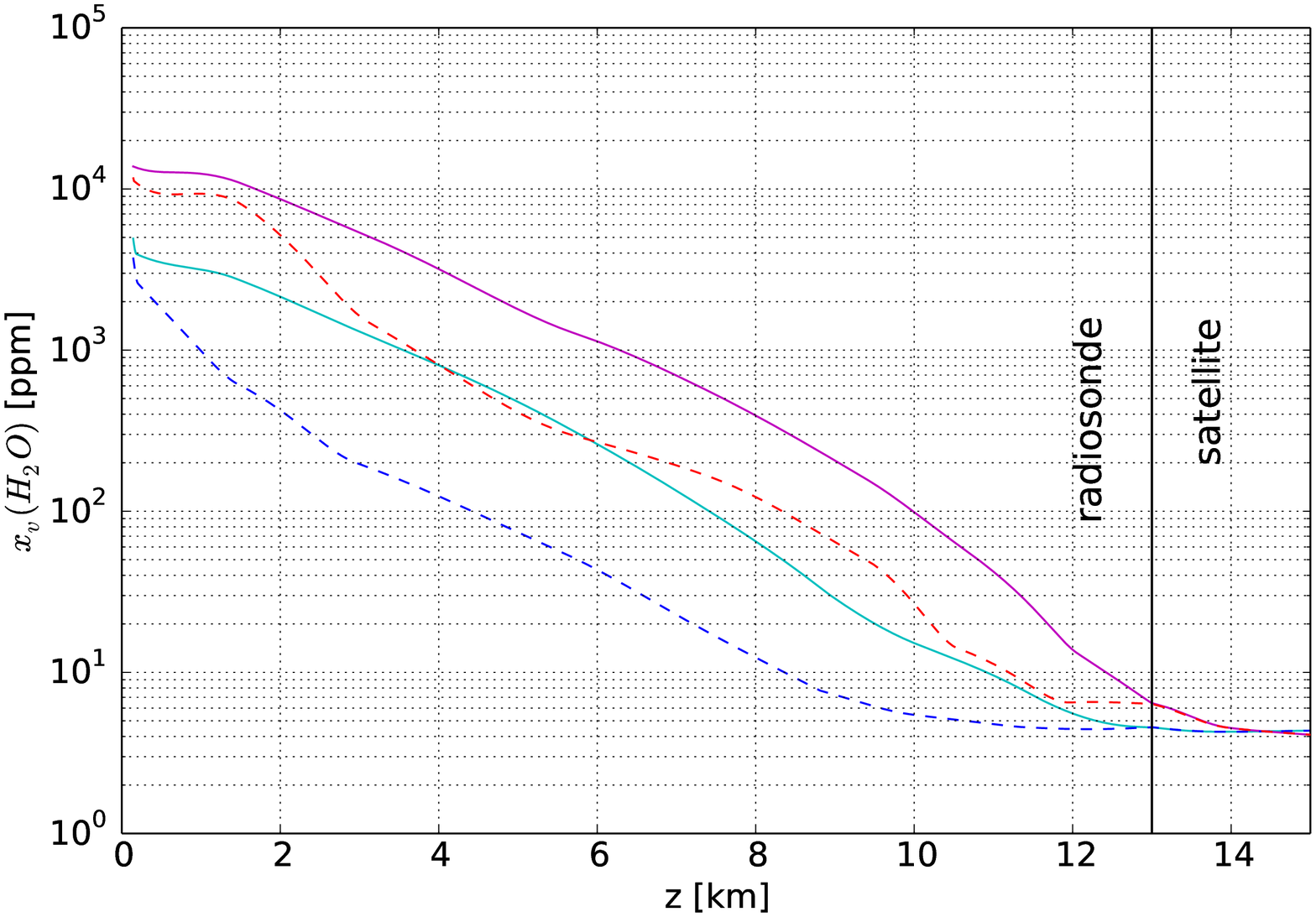}
\caption{({\it Left}) Monthly-average relative humidity profiles
from the Legionowo IGRA station in January (triangles) and July
(squares) for the case of unscreened data (solid), and for the
subset of 10\% driest conditions (dashed) as indicated by $\pH=0.1$ in the
plot legend.  ({\it Right}) Corresponding ${\rm H_2O}$ volume
mixing ratio profiles.  The vertical line indicates transition
altitude above which the satellite data is used. The satellite
water abundances are not subject to the selection criterion. }
\label{fig:RHprofile}
\end{figure*}
Of all trace atmospheric gases, WV is of the greatest relevance for
the cm-wavelength observations.  WV volume mixing ratio (VMR) --- the
number density of ${\rm H_2O}$ molecules relative to number density of
all other species in the atmosphere --- is highest in the
troposphere. At these altitudes satellite data are no longer useful,
and WV mixing ratios or alternatively the relative humidity (RH) for a
given pressure and temperature, must be estimated from radio sounding
data.
The data is publicly available from the Integrated Global Radiosonde
Archive
\footnote{\protect\url{http://www1.ncdc.noaa.gov/pub/data/igra}}
\citep{IGRA_durre_overview_2006} updated on daily basis.
There are several hundreds of
stations around the globe, allowing vertical profiles reconstruction
with the land surface spatial resolution of the order a few hundred 
kilometres, on average.  A typical radiosonde accuracy to perform the
temperature, pressure and RH measurement is about $\degC{0.5}$, $1$
hPa, and $1\%$ respectively \citep{peixoto_climatology_1996}.

For the purpose of our analysis we choose the radiosonde data from
the Legionowo station located near Warsaw (Poland) as they have been archived since
1957.  The data contain daily records of geopotential
height ($z_g$), temperature ($T$), dew point depression ($\Delta
T_d\equiv T-T_d$, where $T_d$ is dew point temperature) wind direction,
and wind speed, probed twice a day (midnight and noon) at fixed
pressure levels ($P_i$), as the radiosonde travels up the atmosphere.
We will hereafter refer to this data as IGRA.  For any given month we use
the daily daytime records between and including the years 2000 and
2014.  We reject the data that have incomplete or incorrect height,
temperature or dew point depression information.  This leaves a few
thousand records per month for further analysis.

At each pressure level and for each $\{T,P,T_d\}$ tuple we derive the
corresponding relative humidity by analytically solving the following equation:
\begin{equation}
T_d(T,\mathrm{RH}) = \frac{\lambda \Gamma(T,\mathrm{RH})}{\beta - \Gamma(T,\mathrm{RH})}
\end{equation}
where 
\begin{equation}
\Gamma(T,\mathrm{RH}) =  \ln\Bigl(\frac{\mathrm{RH}}{100}\Bigr) +  \frac{\beta T}{\lambda+T}
\label{eq:GammaT}
\end{equation}
and $T$ is in Celsius and RH is expressed as percentages (see
e.g. \cite{Lawrence2005} for the derivation).  In Eq.~\ref{eq:GammaT}
the constants are:
$\beta=17.62$ and
$\lambda=\degC{243.12}$ 
\citep{Sonntag1990}. These values yield $T_d$ consistent
with a more rigorous derivation of \cite{Hardy1998} to within
$\degC{0.3}$ for the temperature range from $\degC{-50}$ to
$\degC{50}$, and within the measured range of relative humidities.  See
also \cite{Alduchov1997} to review the values of the constants from
other studies.

We associate the RHs with altitudes using the reconstructed $P(z)$
relation (Sec.~\ref{sec:CIRA}), and employ Akima interpolation
\citep{Akima1970} to create a tabulated version of the profiles.  We
use the sounding data up to $z\sim 13$ km and satellite data at higher
altitudes (Fig.~\ref{fig:RHprofile}).  The near-ground tail of each RH
profile is supplemented with the average RH data point recorded by the
local meteorological station.  This improves adjustment to the RT32
site conditions.

In order to create a low-humidity subset of the data -- conditions
that typically correspond to a cloudless sky -- we select the IGRA
records according to the quantile function $Q(\pH)$ of the RH record
distribution, in each of the pressure levels independently
(sec.~\ref{sec:parametrization}). Thus $\pH$ is a selection parameter,
corresponding to the probability that a random observation at any
pressure level will have the RH value smaller than $Q(\pH)$.  In order to
visualise the selection effect we arbitrarily choose $\pH=0.1$,
corresponding to the 10\% driest conditions for the considered month
in the year-to-year data samples (Fig.~\ref{fig:RHprofile}, dashed
lines).  We will later fine tune this choice using external data.
The resulting yearly samples are averaged into a single profile that
approximates the local climate for a given month.

We convert the reconstructed RH profiles into $\rm{H_2O}$ VMR profiles
($\xv$) using the following formula:
\begin{equation}
\xv = \Bigl(\frac{\rm RH}{100}\Bigr) \frac{P_{\rm sat}(T)}{P}
\label{eq:RH2vmr}
\end{equation}
where $P$ is the total atmospheric pressure, and $P_{\rm sat}$ is the
saturation pressure of water vapour at temperature $T$. $P_{\rm sat}$
is calculated using equation number 10 of \cite{Murphy2005},
reported to be valid within the range of temperatures considered in
this paper.  The reconstructed $\xv$ profiles are shown in
Fig.~\ref{fig:RHprofile} (right panel). At higher altitudes the
profiles are reconstructed using satellite data.

A comparison of CIRA86 and Legionowo radiosonde data in terms of
temperature shows that the two agree very well in the region of the
upper troposphere (Fig.~\ref{fig:cira86_temperature}).  However, there
is a systematical dependence of the upper tropospheric RHs recorded by
radiosondes in the period from 1979 to 1991, on geographical location.
The dependence results from the type of instrumentation that has been
used \citep{soden_assessment_1996}.  For this reason, in this
analysis, we refrain from using data from the previous century, and
rely on the data from the previous and current decades, as technology
exchange it is likely to have mitigated these inconsistencies.
\cite{peixoto_climatology_1996} has also investigated radiosonde
humidity data (recorded in the period from 1973 to 1988), performed
cross-checks with SAGE satellite measurements, and found similar
discrepancies as those pointed out by \cite{soden_assessment_1996},
but confirmed that in the lower and middle troposphere the
discrepancies are not important.

\begin{figure}
\centering
\includegraphics[width=0.48\textwidth]{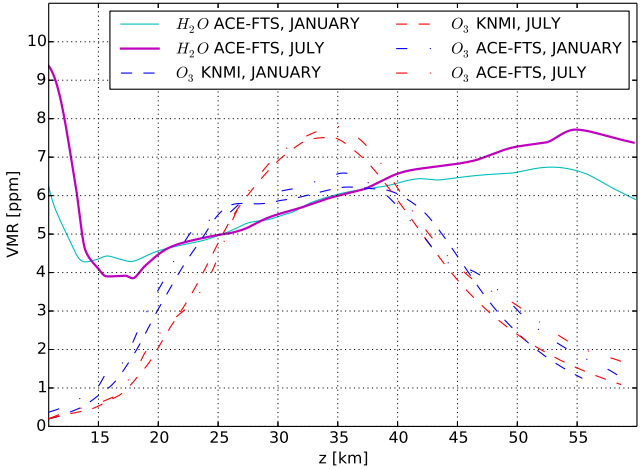}
\caption{WV volume mixing ratio profiles (solid) interpolated at
$\theta_T=53.1\,^\circ$N from ACE-FTS data
in January (cyan,thin) and July (magenta,thick).  The dashed
(dash-dotted) lines represent the ozone VMR profiles interpolated
from the KNMI (ACE-FTS) data in January
(blue) and July (red) (see Sec.~\ref{sec:ozone}).  }
\label{fig:ACEh2o_profile}
\end{figure}

\subsubsection{Water Vapour mixing ratio profile}
\label{sec:WVRprofiles}
We use the data 
from the Fourier Transform Spectrometer 
\citep{ACE-FTS-acp-12-5207-2012} of the Atmospheric Chemistry
Experiment\footnote{The Atmospheric Chemistry Experiment
(ACE), also known as SCISAT, is a Canadian-led mission mainly
supported by the Canadian Space Agency.} 
(ACE-FTS) \citep{Bernath2005}
in order to include the stratospheric water vapour.

The interpolated profile is shown in Fig.~\ref{fig:ACEh2o_profile}.
We combine it with the tropospheric
radiosonde profile using a
linear transformation from one profile to another within the range of
overlapping altitudes (typically $10 \lesssim z \lesssim 13$ km).
We verified, however, that the impact of the stratospheric WV 
is negligible and could be ignored in the case of ground-level sites.

\subsection{Ozone}
\label{sec:ozone}

We model the local atmospheric ozone VMR profile using a combination
of ozonesonde and satellite
data\footnote{\burl{http://www.knmi.nl/research/climate\_chemistry/Data/FKClimatology/}}
provided by the Royal Netherlands Meteorological Institute (hereafter
KNMI) The KNMI ozone data are zonally averaged VMRs measured over 17
different latitudes from $-80\,^\circ$S to $80\,^\circ$N, and at 19
different pressure levels from 0.3 hPa to 1000 hPa.  The data is
compiled from 30 ozone stations and SBUV-SBUV/2 satellite
observations, collected during the years 1980-1991\citep{Paul1998}.

Currently, there is a wealth of ozone data from many satellites and
from radiosondes
(e.g. \cite{ozoneDatabase,fioletov_global_2002,ACE-FTS-acp-12-5207-2012})
that continuously monitor the ozone layer.  In the current work, our
main focus is on the cm-wavelengths, where the ozone impact is
sub-dominant, but for the sake of completeness and accounting for
possible extensions into higher frequencies, we include one of the
available ozone datasets (KNMI) into the analysis.
The
consistency between the profiles obtained form various ozone experiments is shown 
in Fig.~\ref{fig:ACEh2o_profile}.

\subsection{Ground-level meteorological data}
\label{sec:meteoData}
We improve the adjustment of the atmospheric model to the local
conditions at the ground-level by including
data recorded by the IRDAM
WST7000 meteorological station, installed on the roof of
TCfA, near RT32.  The data include
temperatures, pressures, and relative humidities recorded at a high
time resolution.  We used the data covering a
period that slightly exceeds 4 years i.e. from October 2010 to
November 2014. In each month they contribute as a single data point in the RH
profile (Fig.~\ref{fig:RHprofile}) at the altitude of TCfA, but also
influence higher altitudes via smooth interpolation.  We detect a
systematical effect in this RH dataset, resulting from a progressive
deterioration of the humidity sensor over time. The effect leads to an
overestimation of the local RHs by $< 4$\% in the years 2011 and
2014 as compared to the values from 2012, and less than that in the
year 2013.\footnote{In the end of 2011 the station was renewed, hence in
2012 it provided the least biased readouts.
Therefore, the previous and the following years are biassed to a
greater extent.}  In our analysis we ignore these
systematical effects as they are thought to be unimportant.

\subsection{Data selection criteria and model parametrisation}
\label{sec:parametrization}

We are interested in deriving a localised, mean $\Tatm$ and $\tatm$ as
a function of month in clear sky conditions.  Using an unfiltered
data, discussed in Sec.~\ref{sec:IGRA} and Sec.~\ref{sec:meteoData},
would lead to biased results because only a fraction of the data is
obtained at the times when there is no cloud cover.  The fraction of
time with clear sky conditions, depends on the location and the
time of year.  In particular, in the lowlands surrounding the RT32
observing site the fraction of days with the mean daily cloud cover
$\leq 20$\% ranges from $\sim 9\%$ to $\sim 11\%$, on average
\citep{Wos2010}.

As discussed in Sec.~\ref{sec:intro}, there is an anti-correlation
between RH and solar irradiance, which could statistically hint on
clear sky conditions, if one selects data by low, ground-level RHs.
Observations indicate however, that during the cloudiest months such
selection criterion returns a considerable false-positive rates, as
discussed in Sec.~\ref{sec:clear_sky_detection}.  However, there is
another complication regarding this selection scheme.  The correlation
between water vapour content at the ground level and that at higher
altitudes is rather weak, or non-existent.  Some correlation exists
only between the neighbouring pressure levels in the lower
troposphere.  For this reason, the parametrisation of the selection
criterion by low RHs at a single, fixed pressure level lacks a very
desired feature: that the mean column PWV should decrease
monotonically as one straightens the selection criterion.

Since we are interested in an assessment of the brightness temperature
in the conditions that are {\em statistically} compatible with a
clear sky situation in terms of the column PWV content, we
will filter the data coherently at all pressure levels.
For a given month and year, we use only
the IGRA records that yield
\begin{equation}
\label{eq:selection}
{\rm RH}<Q_{{\rm RH},i,j}(\pH),
\end{equation} 
where $Q_{{\rm RH},i,j}(\pH)$ is the quantile function of the RH
distribution at $i$'th pressure level in $j$'th month.  We use the
same $\pH$ parameter consistently for every pressure level, month, and
for the ground-based meteorological data.  The resulting vertical
profile will be, in most cases, a combination of data taken at different
days, but the parametrisation assures that imposing a stronger
selection criterion (i.e. lowering $\pH$) results in a lower column
PWV, as expected.  In order to average over the year-to-year
variability of PWV,
and approximate the climate for a given month, we split the
IGRA records into yearly sub-samples, which we analyse separately.
Then, for any given month and pressure level the mean RH
is calculated for all years.
The exact value of the $\pH$
selection parameter, that would be compatible with cloudless skies, and
would not concern only the driest conditions, is unknown, but can be
constrained by independent PWV measurements, performed in  clear
sky conditions (see Sec.~\ref{sec:PWVdata}).

The downside of such parametrisation is that the probability
distribution function (PDF) for the $\pH$ selection parameter, by
construction, is zero outside of $[0,1]$ range. This is because $\pH$
is associated with the probability of obtaining a RH measurement
smaller than the quantile function (Eq.~\ref{eq:selection}) for a
given pressure level, month and year.  The final monthly-average
profile is a multi-year mean, therefore even for the maximum value of $\pH=1$ --
i.e. when no data is filtered out -- the resulting average model will
not be able to describe an individual day with PWV values
above the mean. However, in the current
approach we are focused on a parametrisation that is suitable
for the average clear sky model, and will accept these limitations
since we are not going to analyse the significance of deviations of
individual measurements from the mean.

In order to account for the large diurnal variation of RH and to enable
a meaningful comparison between different days, we select the RH data
records acquired between hours 10 and 14 (UTC+1), where the
temperatures should be least affected by the diurnal variation of the
solar irradiance. Hence, the differences in RH between different days, better
reflect the actual PWV content variations, and not temperature
variations.

\subsection{PWV measurements}
\label{sec:PWVdata}

We have been monitoring PWV, expressed in column millimetres of water
($w$), in the clear sky conditions since June 2013.  A compilation from
the first data release
is summarised in Table~\ref{tab:PWV}.
The measurements
were performed using a hand-held MICROTOPS II sun photometer at the
Meteorological Observatory of the Department of Meteorology and
Climatology of the Nicolaus Copernicus University in Toru\'n (hereafter DMC).
In what follows, we will refer to our sun photometer
PWV measurements as {\trpwv}.

The atmospheric column PWV is calculated along LOS towards the Sun,
based on photometric measurements of water absorption peak at 936 nm,
and of the continuum at 1020 nm (without water absorption).  The PWV
resulting from Beer-Lambert-Bouguer's law is converted to the zenith
column water abundance based on the Sun's zenith distance at the
time and location of the measurement. The accuracy of the measurement is 0.1 mm.
Using a reference MICROTOPS II sun photometer we verified that with
the factory settings the systematical differences to measure PWV
between different instruments is $\Delta w \lesssim 0.1$ mm.

The data have been collected only in clear sky
conditions, which allows us to quantify the PWV variations due to air
masses carrying different amounts of water vapour,
depending on seasons and winds.
The data are typically taken every hour and
every observation contains several measurements that are
averaged.  In stable clear sky conditions the differences between
hourly samples are small (relative to our measurement precision) and
we use a daily mean as a single data record. Then, the monthly
average, standard deviation and extreme values are calculated from
daily records (Table~\ref{tab:PWV}).

\begin{figure*}
\centering
\includegraphics[width=\textwidth]{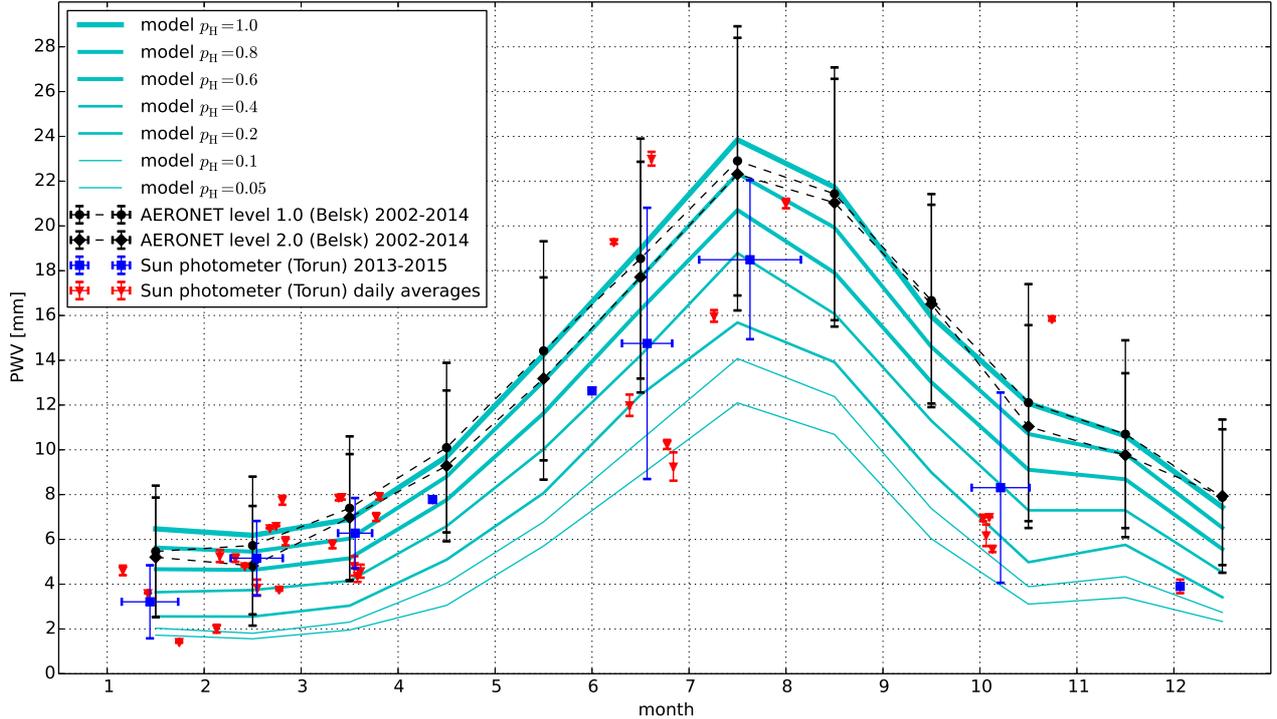}
\caption{A compilation of PWV measurements.  The monthly averages from
our {\trpwv} sample are marked with blue squares while the daily
averages are marked with red triangles. The error bars correspond to
$\pm 1\sigma$ deviation from the mean. The black diamonds (dots) are
the AERONET data level 2.0 (1.0) from the Belsk station. The cyan lines
represent our PWV models reconstructed from climatology data
described in Sec.~\ref{sec:data}. The thickness of an individual
solid line increases with $\pH$. The increases correspond to a
progressively less aggressive selection of the climatology data by
low RH values (Sec.~\ref{sec:parametrization}).  }
\label{fig:PWV}
\end{figure*}

We also utilise the publicly available PWV data from 
AERONET\footnote{\url{http://aeronet.gsfc.nasa.gov}} robotic stations,
which automatically trace the Sun and measure the atmospheric water lines
spectrum. We use the data from the Belsk station in Poland (near Warsaw),
collected between (and including) years 2002 and 2014.  Since the
AERONET data are collected automatically and exclude only the periods
of rain, the raw data may be contaminated by the presence of clouds
and therefore should be more compatible with the average PWV
abundances reconstructed from climatology data: i.e. without imposing
any selection criteria.  The AERONET data are essentially provided at three
levels of post-processing. Level 1.0 is raw data
(Fig.~\ref{fig:PWV}).  Level 1.5 data are automatically
screened for clouds based on number of simple criteria involving time
and spectral stability of the measured optical depths
\citep{Smirnov2000}, and level 2.0 data, which we use in our analysis, are
further corrected manually for glitches and any other abnormalities
that could elude automatic screening.

The difference between our measurements and the AERONET level 2.0 data
(Fig.~\ref{fig:PWV}) is that the latter have clouds removed directly
from the LOS towards the Sun, thus not guaranteeing a cloudless sky.
Our sample however, in most cases was taken during cloudless days, at
the penalty of fewer measurements.

\section{Solving radiative transfer through atmospheric layers}
\label{sec:transfer}

In this section we briefly review the principal equations of 
radiative transfer and outline the calculation scheme.

Based on a model of the vertical structure of the atmosphere we calculate
the column PWV ($w$), atmospheric brightness temperature ($\Tatm$), 
optical depth ($\tau$) and the corresponding transmittance ($\tatm=e^{-\tau}$)
using the AM program (version 7.2) -- a
publicly-available radiative transfer solver, developed at the
Harvard-Smithsonian Center for Astrophysics \citep{AM}.

Propagation of radiation through a medium is given by the
radiative transfer equation
\begin{equation}
\frac{\d I_\nu}{\d s} = -\kappa_\nu I_\nu + \epsilon_\nu,
\label{eq:radiativeTransfer}
\end{equation}
where $I_\nu$ is the specific intensity and
$\kappa_\nu = -\frac{\d\tau_\nu}{\d s}$,
is the total opacity due to absorption and scattering,
$\tau_\nu$ is the optical depth, and $\epsilon_\nu$ is the emissivity
along the propagation path $s$.
In the case of local
thermodynamic equilibrium (LTE) the radiative transfer equation can be
rewritten ~\citep{Wilson2009} as:
\begin{equation}
-\frac{1}{\kappa_\nu}\frac{dI_\nu}{ds} = \frac{dI_\nu}{d\tau} = I_\nu - B_\nu(T),
\label{eq:radiativeTransfer2}
\end{equation}
where $B_\nu(T)$ is the Planck radiance.
The solution is given by:
\begin{equation}
I_\nu(s) = I_\nu(0) e^{-\tau_\nu(s) }  + \int_0^{\tau(s)} B_\nu (T(\tau))  e^{-\tau_\nu }d\tau.
\label{eq:radiativeTransfer2int}
\end{equation}
For a thin atmospheric layer, inside which the temperature can be assumed
constant, Eq.~\ref{eq:radiativeTransfer2int} can be integrated as
\begin{equation}
I_\nu(s) = I_\nu(0) e^{-\tau_\nu(s) }  + B_\nu (T) (1- e^{-\tau_\nu(s)}),  
\label{eq:radiativeTransferLTE}
\end{equation}
where the first term on the right hand side is the incident radiance
that is exponentially attenuated along the propagation path.  The
second term amounts to the thermal emission of the layer corrected for
the self-absorption, required to maintain the assumed LTE.
The spectrum of the optical
depth is medium dependent and can be
derived from the quantum-mechanical properties of atoms and molecules
present in the medium.  The effects leading to a violation of the LTE,
such as thermal conduction or convection are reasonably neglected.

The radiative transfer equation is solved 
for each absorbing (emitting) species and for each  
atmospheric layer. We use
$N_L=300$ stacked layers, each characterised by its pressure ($P$),
temperature ($T$), geometrical thickness, and chemical composition: a
mixture of nitrogen, oxygen, ozone, and water vapour defined in terms
of VMRs.  The layers are homogeneously distributed in $\log$-altitude
space, which roughly corresponds to a homogeneous distribution in
pressure space.  We use the same definition of layers for all
atmospheric species.  The lowermost layer altitude, $z=133$ m, is
chosen to coincide with the altitude of RT32.  The uppermost layer
altitude is assumed to be $z_{\rm max}=60$ km.

It is assumed that the incident (background) radiation has initially
Planckian distribution ($B_\nu(T)$) with $T=T_{\rm CMB}=2.726$
\citep{Fixsen2009} -- the thermodynamic temperature of the Cosmic
Microwave Background radiation (CMB). Thus, our definition of $\Tatm$
includes the contribution of CMB.

Calculations are performed towards zenith or at directions located
at the zenith distance $z_d$, and layers are assumed to have a flat
geometry.  
The brightness temperature $\Tatm$ is
obtained by solving 
\begin{equation}
\Iatm = B_\nu(\Tatm)
\label{eq:Iatm2Tatm}
\end{equation}
where $\Iatm$ is the output radiance at the
lowermost atmospheric layer.
The brightness temperature spectrum is calculated within the frequency
range $\nu=[1, 60]$ GHz with $50$ MHz resolution.  Details of the
physical processes taken into account are described in the AM
program technical memo \citep{AM}.

\section{Results}
\label{sec:results}

\subsection{Precipitable Water Vapour}
\label{sec:PWV}

\begin{figure}
\centering
\includegraphics[width=0.45\textwidth]{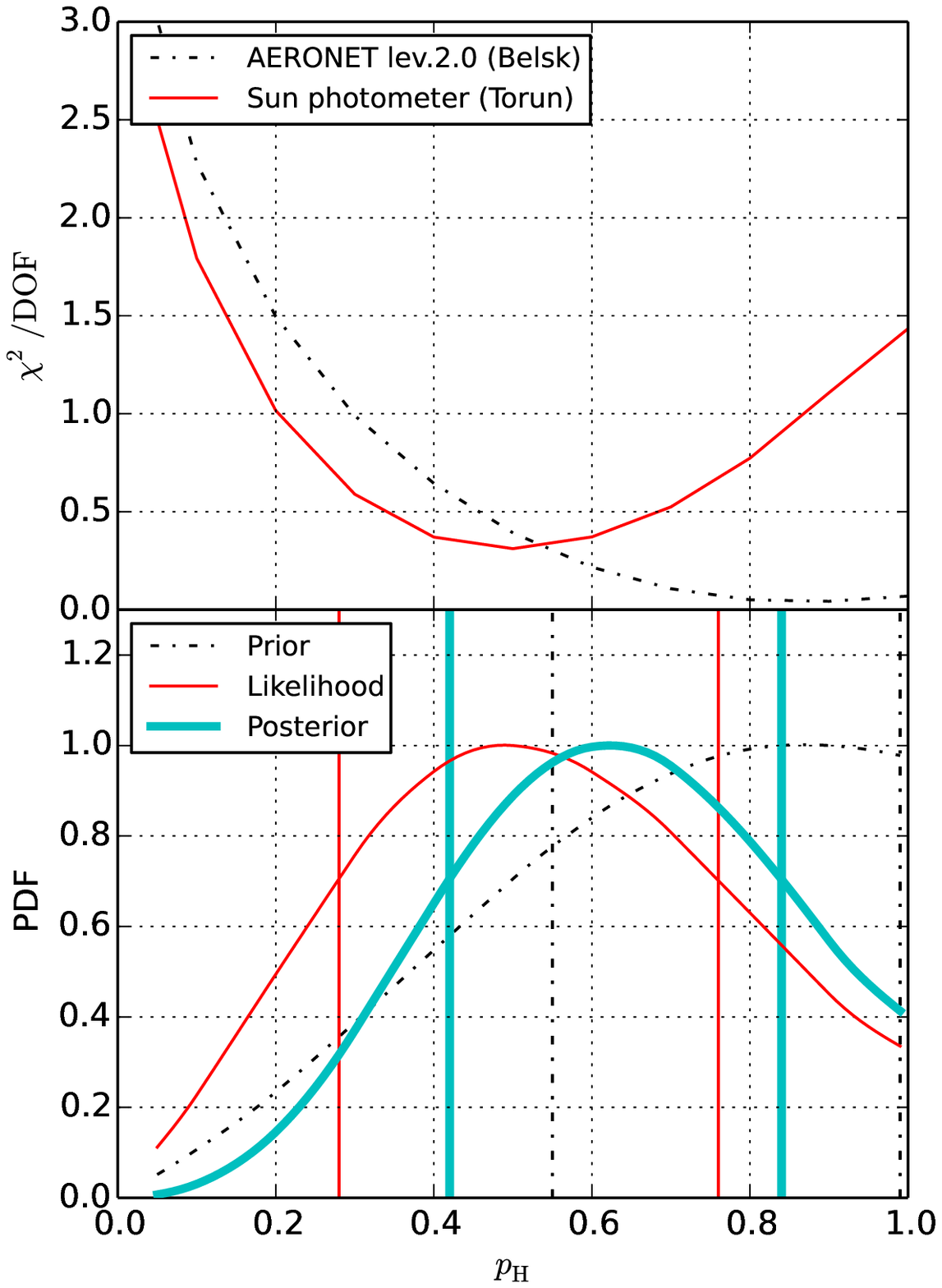}
\caption{Constraints on $\pH$ selection parameter (see
Sec.~\ref{sec:parametrization}).  The vertical lines in the bottom
panel encompass the 68\% CR, as summarised in
Table~\ref{tab:PWVchisq}. The three PDFs are obtained from the
AERONET/Belsk PWV data (black, dash-dotted),
Toru\'n MICROTOPS measurements (red, solid thin),
and the two datasets combined (solid thick).  }
\label{fig:PWVchisq}
\end{figure}

A compilation of the {\trpwv} measurements (Sec.~\ref{sec:PWVdata}) is
shown in Fig.~\ref{fig:PWV}, and the corresponding data are gathered
in Table~\ref{tab:PWV}. In this figure solid (cyan) lines represent
the PWV models calculated using the climatological data
discussed in Sec.~\ref{sec:CIRA}--~\ref{sec:meteoData}. The line width
is increased with the increases of the $\pH$ selection
parameter  (Sec.~\ref{sec:parametrization}).  For any given month,
the reconstructed column PWV is assumed to occur at mid-month.

We constrain the value of the $\pH$ parameter, which implies the PWV 
level that is expected to occur in clear sky conditions.
We use a $\chi^2$ minimisation, neglecting the cross-month covariance:
\begin{equation}
\chi^2= \sum_{i=1}^{12}\left[\langle w\rangle_i - \mathcal{M}_i(\pH)\right]^2 / \sigma_i^2,
\label{eq:chisq}
\end{equation}
where $\langle w\rangle_i$ and $\sigma_i^2$ are the $i$'th month PWV
mean and standard deviation respectively, and $\mathcal{M}_i(\pH)$ is
the model PWV value interpolated at the locus of the $i$'th month data
point.  When the monthly variance estimate is unknown (due to missing
data), it is interpolated from the neighbouring months (with
continuity across Dec/Jan).

Given the PWV data $\mathcal D$, and the reconstructed PWV model
$\mathcal M$, parametrised by $\pH$, we define the posterior
probability for $\pH$ using the Bayes theorem:
\begin{equation}
\mathcal P(\pH | \mathcal M,\mathcal D)  \propto  \mathcal L(\mathcal D | \mathcal M,\pH) \Pi(\pH|\mathcal M)
\label{eq:PWVbayes}
\end{equation}
where $\mathcal L(\mathcal D | \mathcal M,\pH)$ is the likelihood of
the data given the model and $\Pi(\pH|\mathcal M)$ is the prior
imposed on the parameter probability distribution function (PDF).  By
design, the PDF for the $\pH$ parameter is zero outside $[0,1]$
range (see Sec.~\ref{sec:parametrization}).  We assume the AERONET
data as a prior and obtain the maximum posterior constraint on the
$\pH$ parameter and calculate the 68\% confidence interval. The
likelihood function is probed using the sun photometer
data (Fig.~\ref{fig:PWVchisq}).
However, we also analyse each of the datasets alone. Constraints on
the $\pH$ parameter are gathered in Table~\ref{tab:PWVchisq}.  As
indicated by the vertical lines in Fig.~\ref{fig:PWVchisq}, {\trpwv}
and the AERONET/Belsk datasets are compatible at the 68\% CL, but the
maximum-likelihood $\pH$ parameter value is lower for the earlier
data, which we attribute to the fact that the {\trpwv} sample was
collected under excellent weather conditions, which guaranteed a
cloudless sky, and therefore statistically favoured lower PWV
abundances.

From Fig.~\ref{fig:PWV} it is clear that
the year-around distribution of AERONET/Belsk PWV
follows very closely the shape of the
reconstructed models with high $\pH$ values (corresponding to weak
data selection). The strength of the correlation is also reflected in
the smallness of the best fit $\chi^2/{\rm DOF}$ value, although the
dispersion of the individual PWV measurements (from which the monthly
variance is calculated) is relatively large in both data samples.
Clearly, AERONET/Belsk level 1.0 dataset corresponds to a greater $\pH$
value than level 2.0 dataset, as expected.

\begin{table}
\caption{ Compilation of column PWV measurements ({\trpwv}) performed in 
clear sky conditions using MICROTOPS II sun photometer
(Sec.~\ref{sec:PWVdata}).  }
\Beginruledtabular
\begin{tabular}{rcccc}
\BeginruledtabularSec
Location\footnotemarkTAB{$^a$}&\multicolumn{4}{l}{$18^\circ\, 34'\, 04.8''\, \mathrm{E},\, 53^\circ\, 01'\, 12.0''\, \mathrm{N}$}\\
Period &\multicolumn{4}{l}{2013/06/07 -- 2015/04/13}\\
Duration\footnotemarkTAB{$^b$} &\multicolumn{4}{l}{$\sim 17$ months}\\
\hline

Month  & 
$\langle w \rangle$\footnotemarkTAB{$^c$}  & 
Min/Max\footnotemarkTAB{$^d$}  & 
\# of days\footnotemarkTAB{$^e$} & 
Cmt\footnotemarkTAB{$^f$}\\
& 
[mm]& 
[mm] & 
& 
\\
$ 1 $    &   $ 3.2 \pm 1.6 $   &   $ 1.4 \, / \, 4.6 $   &  $ 3 $   &   \\
$ 2 $    &   $ 5.2 \pm 1.7 $   &   $ 2.0 \, / \, 7.7 $   &  $ 10 $   &   \\
$ 3 $    &   $ 6.3 \pm 1.6 $   &   $ 4.4 \, / \, 7.9 $   &  $ 8 $   &   \\
$ 4 $    &   $ 7.8 \pm 0.1 $   &   $ 7.8 \, / \, 7.8 $   &  $ 1 $   &  SD \\
$ 5 $    &   $ 12.6 \pm 0.1 $   &   $ 12.6 \, / \, 12.6 $   &  $ 1 $   &  SD \\
$ 6 $    &   $ 14.8 \pm 6.1 $   &   $ 9.3 \, / \, 23.0 $   &  $ 5 $   &   \\
$ 7 $    &   $ 18.5 \pm 3.5 $   &   $ 16.0 \, / \, 21.0 $   &  $ 2 $   &   \\
$ 8 $     &    &    &   $ 0 $   &  ND  \\
$ 9 $     &    &    &   $ 0 $   &  ND  \\
$ 10 $    &   $ 8.3 \pm 4.2 $   &   $ 5.6 \, / \, 15.8 $   &  $ 5 $   &   \\
$ 11 $     &    &    &   $ 0 $   &  ND  \\
$ 12 $    &   $ 3.9 \pm 0.1 $   &   $ 3.9 \, / \, 3.9 $   &  $ 1 $   &  SD \\

\EndruledtabularSec
\end{tabular}
\Endruledtabular

\footnotetextTAB{$^a$Geodetic coordinates of the observation site (DMC). }
\footnotetextTAB{$^b$Effective number of observed months. No-observation periods: Aug--Sep 2013 and Jul--Oct 2014.}
\footnotetextTAB{$^c$Monthly average and a standard deviation of PWV, both calculated using daily means. }
\footnotetextTAB{$^d$PWV data extremal values}
\footnotetextTAB{$^e$The total number of days observed in a given month.}
\footnotetextTAB{$^f$Comment: SD - single day mean and standard deviation, ND - no data}
\label{tab:PWV}
\end{table}

\begin{table}
\caption{
68\% CL constraints on $\pH$ selection parameter from the sun
photometer data (Table~\ref{tab:PWV}), from the AERONET/Belsk data and
from the two datasets combined.
}
\Beginruledtabular
\begin{tabular}{rccc}
\BeginruledtabularSec
& Sun Phot. & AERONET  & Combined \\
&  (Toru\'n) & (Belsk, lev. 2.0) &  \\
$\min\bigl(\frac{\chi^2}{{\rm DOF}}\bigr)$ & 
$\sim 0.31$ & 
$\sim 0.04$ & 
NA\\
$\pH$  & 
(0.28,0.76) & 
(0.55,1.00) & 
(0.42,0.84) \\
MOD$(\pH)$ & 
$ 0.49^{+0.27}_{-0.21} $ &
$ 0.87^{+0.12}_{-0.32} $ & 
$ 0.62^{+0.22}_{-0.20} $ \\
${\rm EX}(\pH)$ &
0.53 & 
0.65 & 
0.62 \\
\EndruledtabularSec
\end{tabular}
\Endruledtabular
\label{tab:PWVchisq}
\end{table}

\subsection{Average clear sky atmospheric brightness temperature and optical depth: case for Toru\'n, Poland}
\label{sec:Tatm}

With the reconstructed vertical structure of the atmosphere
(Sec.~\ref{sec:data}) we now determine the local $\Tatm$ and $\tau$
using the radiative transfer solver described in
Sec.~\ref{sec:transfer}.

\begin{figure}
\centering
\includegraphics[width=0.48\textwidth]{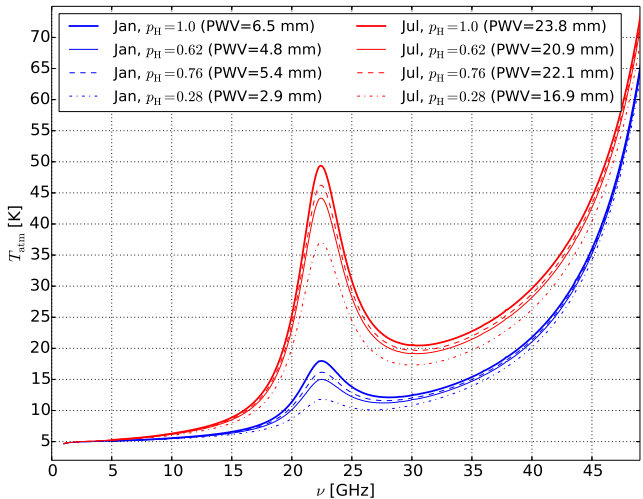}
\caption{Average atmospheric brightness temperature as a function of
frequency. The upper (red) lines correspond to July and lower (blue)
lines correspond to January in Toru\'n. Within each group the thick
solid line traces $\Tatm$ corresponding to the model obtained
without imposing any selection criteria on the climatology data
($\pH=1.0$). Thin solid lines are calculated according to the best
fit models found using the combined (Toru\'n + AERONET/Belsk) PWV
data samples.  The best fit $\pH$ selection parameter values are
given in the plot legend and in Table~\ref{tab:PWVchisq}. The dashed
and dash-dotted lines enclose the 68\% confidence region, derived
using the Toru\'n PWV data sample alone ({\trpwv}).  All curves are
calculated for the zenith distance $z_d=0^\circ$ at the RT32
altitude ($z=0.133$ km).
}
\label{fig:atmTb}
\end{figure}

\begin{table*}
\caption{
68\% CL constraints on the clear sky, zenith, mean atmospheric
brightness temperature ($\Tatm$) and optical depth ($\tau$).  The
values for the selected frequencies and months are calculated
according to the best fit PWV model selected by the combined
AERONET/Belsk and Toru\'n data samples (Table~\ref{tab:PWVchisq}).  The
asterisk (*) marks the errors that are rounded up to appear non-zero at
the provided accuracy.  $A$ and $B$ are the
linear-fit constants of the $\Tatm=A \langle w\rangle + B$ and
$\tatm=A \langle w\rangle + B$ scaling relations.  $A$ and $B$ are
written in scientific notation with the decimal exponent in
parentheses.
}
\Beginruledtabular

\begin{tabular}{rccccccccc}
\BeginruledtabularSec
& $\nu$ [GHz]  &  5  & 15  & 22  & 30 &  5  & 15  & 22  & 30 \\\hline
Month & $\langle w\rangle$     & \multicolumn{4}{c}{$\Tatm$ } & \multicolumn{4}{c}{$\tau$ }\\
&   [mm]                   &    [K]  &    [K] &    [K] &    [K] &   $\times 10^{-3}$ & $\times 10^{-2}$  & $\times 10^{-2}$  & $\times 10^{-2}$  \\
1   &   $  4.8^{+1.1}_{-1.0}  $   &   $  5.10^{+0.01}_{-0.01}  $   &   $  6.4^{+0.1}_{-0.1}  $   &   $  14.6^{+1.9}_{-1.7}  $   &   $  11.7^{+0.5}_{-0.5}  $   &   $  9.50^{+0.03}_{-0.03}  $   &   $  1.5^{+0.1}_{-0.1}  $   &   $  4.7^{+0.8}_{-0.7}  $   &   $  3.6^{+0.2}_{-0.2}  $   \\
2   &   $  4.7^{+0.9}_{-0.9}  $   &   $  5.09^{+0.01}_{-0.01}  $   &   $  6.3^{+0.1}_{-0.1}  $   &   $  14.4^{+1.5}_{-1.5}  $   &   $  11.7^{+0.4}_{-0.4}  $   &   $  9.48^{+0.03}_{-0.04}  $   &   $  1.4^{+0.1*}_{-0.1*}  $   &   $  4.6^{+0.6}_{-0.6}  $   &   $  3.6^{+0.2}_{-0.2}  $   \\
3   &   $  5.3^{+1.0}_{-1.0}  $   &   $  5.09^{+0.01}_{-0.01}  $   &   $  6.4^{+0.1}_{-0.1}  $   &   $  15.5^{+1.7}_{-1.7}  $   &   $  11.9^{+0.5}_{-0.5}  $   &   $  9.39^{+0.04}_{-0.04}  $   &   $  1.5^{+0.1}_{-0.1*}  $   &   $  5.0^{+0.7}_{-0.7}  $   &   $  3.6^{+0.2}_{-0.2}  $   \\
4   &   $  7.8^{+1.2}_{-1.2}  $   &   $  5.12^{+0.01}_{-0.01}  $   &   $  6.7^{+0.2}_{-0.1}  $   &   $  19.9^{+2.1}_{-1.9}  $   &   $  13.0^{+0.6}_{-0.5}  $   &   $  9.34^{+0.06}_{-0.06}  $   &   $  1.5^{+0.1}_{-0.1}  $   &   $  6.6^{+0.9}_{-0.8}  $   &   $  4.0^{+0.2}_{-0.2}  $   \\
5   &   $  11.8^{+1.6}_{-1.6}  $   &   $  5.16^{+0.01}_{-0.02}  $   &   $  7.3^{+0.2}_{-0.2}  $   &   $  26.9^{+2.8}_{-2.8}  $   &   $  14.9^{+0.7}_{-0.7}  $   &   $  9.36^{+0.06}_{-0.07}  $   &   $  1.7^{+0.1}_{-0.1}  $   &   $  9.3^{+1.2}_{-1.1}  $   &   $  4.6^{+0.3}_{-0.3}  $   \\
6   &   $  16.3^{+1.7}_{-1.9}  $   &   $  5.20^{+0.02}_{-0.02}  $   &   $  7.9^{+0.2}_{-0.2}  $   &   $  34.7^{+3.0}_{-3.3}  $   &   $  17.0^{+0.8}_{-0.9}  $   &   $  9.40^{+0.08}_{-0.07}  $   &   $  1.9^{+0.1}_{-0.1}  $   &   $  12.3^{+1.3}_{-1.4}  $   &   $  5.4^{+0.3}_{-0.3}  $   \\
7   &   $  20.9^{+1.9}_{-1.9}  $   &   $  5.25^{+0.02}_{-0.02}  $   &   $  8.5^{+0.2}_{-0.2}  $   &   $  42.8^{+3.2}_{-3.2}  $   &   $  19.2^{+0.9}_{-0.8}  $   &   $  9.47^{+0.08}_{-0.07}  $   &   $  2.1^{+0.1}_{-0.1}  $   &   $  15.5^{+1.4}_{-1.4}  $   &   $  6.1^{+0.3}_{-0.3}  $   \\
8   &   $  18.1^{+2.4}_{-1.8}  $   &   $  5.23^{+0.02}_{-0.02}  $   &   $  8.1^{+0.3}_{-0.2}  $   &   $  38.0^{+4.0}_{-3.2}  $   &   $  17.9^{+1.1}_{-0.8}  $   &   $  9.43^{+0.08}_{-0.07}  $   &   $  2.0^{+0.1}_{-0.1}  $   &   $  13.6^{+1.7}_{-1.4}  $   &   $  5.7^{+0.4}_{-0.3}  $   \\
9   &   $  13.0^{+1.9}_{-1.7}  $   &   $  5.18^{+0.02}_{-0.02}  $   &   $  7.5^{+0.2}_{-0.2}  $   &   $  29.0^{+3.3}_{-2.9}  $   &   $  15.5^{+0.8}_{-0.8}  $   &   $  9.39^{+0.07}_{-0.07}  $   &   $  1.8^{+0.1}_{-0.1}  $   &   $  10.1^{+1.4}_{-1.2}  $   &   $  4.8^{+0.3}_{-0.3}  $   \\
10   &   $  9.3^{+1.8}_{-1.7}  $   &   $  5.15^{+0.01}_{-0.02}  $   &   $  7.0^{+0.2}_{-0.2}  $   &   $  22.6^{+3.1}_{-3.0}  $   &   $  13.8^{+0.8}_{-0.8}  $   &   $  9.36^{+0.07}_{-0.07}  $   &   $  1.6^{+0.1}_{-0.1}  $   &   $  7.7^{+1.3}_{-1.2}  $   &   $  4.2^{+0.3}_{-0.3}  $   \\
11   &   $  8.7^{+1.3}_{-1.4}  $   &   $  5.13^{+0.01}_{-0.01}  $   &   $  6.9^{+0.2}_{-0.2}  $   &   $  21.3^{+2.2}_{-2.3}  $   &   $  13.5^{+0.6}_{-0.6}  $   &   $  9.40^{+0.06}_{-0.06}  $   &   $  1.6^{+0.1}_{-0.1}  $   &   $  7.2^{+0.9}_{-1.0}  $   &   $  4.2^{+0.2}_{-0.2}  $   \\
12   &   $  5.7^{+1.2}_{-1.0}  $   &   $  5.11^{+0.01}_{-0.01}  $   &   $  6.5^{+0.1}_{-0.1}  $   &   $  16.2^{+2.0}_{-1.8}  $   &   $  12.2^{+0.5}_{-0.5}  $   &   $  9.48^{+0.04}_{-0.05}  $   &   $  1.5^{+0.1}_{-0.1}  $   &   $  5.3^{+0.8}_{-0.7}  $   &   $  3.7^{+0.2}_{-0.2}  $   \\
Mean  &   $  10.5  $  &  $  5.15  $  &  $  7.1  $  &  $  24.7  $  &  $  14.4  $  &  $  9.42  $  &  $  1.7  $  &  $  8.5  $  &  $  4.5  $    \\ 
St.Dev.  &   $  5.5  $  &  $  0.05  $  &  $  0.7  $  &  $  9.7  $  &  $  2.6  $  &  $  0.05  $  &  $  0.2  $  &  $  3.7  $  &  $  0.9  $    \\\hline 
$A$  &     &  $  9.812 (-03)   $   &  $  1.339 (-01)   $   &  $  1.753 (+00)   $   &  $  4.621 (-01)   $   &  $  -4.544 (-06)   $   &  $  4.046 (-04)   $   &  $  6.676 (-03)   $   &  $  1.567 (-03)   $   \\
$B$  &    &  $  5.047 (+00)   $    &  $  5.713 (+00)   $    &  $  6.185 (+00)   $    &  $  9.492 (+00)   $    &  $  9.475 (-03)   $    &  $  1.249 (-02)   $    &  $  1.458 (-02)   $    &  $  2.811 (-02)   $    \\

\EndruledtabularSec
\end{tabular}

\Endruledtabular
\label{tab:seasonsDependence}

\end{table*}

In Fig.~\ref{fig:atmTb} we plot the best-fit local $\Tatm$ models up to
the Q-band frequencies.  The most prominent atmospheric feature is
associated with the water line that is Doppler- and pressure-broadened
about the resonance frequency $\nu_0\approx 22.235$ GHz.  On the top
of the broad band emission a very weak and sharp contribution is found
due to a thermally-excited stimulated emission. This is seen as a tiny
increase of $\Delta\Tatm\approx 0.15$ K exactly at the resonance
frequency $\nu_0$, however effects of fluctuating PWV levels due to
atmospheric turbulence (see Sect.~\ref{sec:TatmPWV}) generate $\Tatm$
variations of the same order in the time scale of hours or
less.

Clearly, $\Tatm$ and $\tau$ are strongly season-dependent, as is the
level of PWV.  The high-frequency tail of $\Tatm$ in
Fig.~\ref{fig:atmTb} is caused by the contribution from the oxygen
$O_2$ line that is insensitive to PWV content.  The only way to
mitigate this emission is to observe at higher altitudes, i.e. through
a thinner atmosphere.  In Fig.~\ref{fig:atmTb} the legend provides the
value of column PWV in millimetres for each model.  For comparison, the mean
PWV level on the South Pole in the austral winter is about 0.26 mm
\citep{Bussmann2005} which is $\sim 25$ times lower as compared to
the mean PWV content in January in Legionowo (Poland) and $\sim 12$
times lower than the mean PWV content in clear sky conditions
measured in Toru\'n (Table~\ref{tab:PWV}). 
However, since the WV brightness temperature spectrum depends
on temperature,
a comparison of radiative properties between the two sites 
based on the PWV levels difference is not straightforward. 

The impact of ozone is seen only in spectral lines, which are weak
in the Q-band,  as compared to the effects caused by
atmospheric instabilities.  We re-calculate the spectra with the
resolution of 5 kHz and observe that the lines amplitude is well below 1 K
above the continuum at the resonance frequencies.
However, the ozone contribution becomes more important at higher
frequencies.

From Fig.~\ref{fig:atmTb} it is clear that lower PWV levels result in
lower $\Tatm$ values. The stratospheric PWV levels, as traced by
satellites, are roughly constant throughout the year.  We observe
that above 13 km the levels are very low, in the order of a few
$\mu$-meters: $w\sim 0.005$ mm in January and July, and therefore the
stratospheric PWV impact on $\Tatm$ seems unimportant for the
ground level sites.

We use the best fit model determined by the combined AERONET/Belsk and
Toru\'n data samples (Table~\ref{tab:PWVchisq}) to calculate the local
$\Tatm$ and $\tau$ for each month.  The result is shown in
Table~\ref{tab:seasonsDependence}.  In that table the error bars
correspond to the variation of the $\pH$ selection parameter within
the 68\% CR.

\subsection{Atmospheric approximations and zenith distance dependence}
\label{sec:approximations}
In common observational practice simplifying approximations are introduced
at the cost of accuracy.  These approximations rely on 
assumptions that the atmosphere:
(i) is composed of a single homogeneous layer,
(ii) has a flat geometry and
(iii) is optically thin.
In the following sections we briefly discuss these assumptions
in the light of the reconstructed atmospheric model. 

\subsubsection{Single-layer atmosphere}
\label{sec:SL}
A single layer atmosphere has a single physical temperature $T$
(Eq.~\ref{eq:radiativeTransferLTE}) that is often
assumed to be somewhere between 250 K and 290 K.
However, given the reconstructed atmospheric model it is easy to
derive this value.

From Eq.~\ref{eq:radiativeTransfer2int}
the atmospheric emission of a multi-layer, flat atmosphere
with a vertical temperature profile can be written as:
\begin{equation}
\Iatm = B_\nu(\Tcmb) e^{-\tau(z_d)} + \sum_{i=1}^N B_\nu(T_i) \bigl( 1-e^{-\tau_i(z_d)}\bigr),
\label{eq:Iatm}
\end{equation}
where $T_i$ is the $i$'th layer temperature, $\tau_i$ is its optical depth,
and N is the total number of layers (Sec.~\ref{sec:transfer}).
$\Iatm$ is the calculated atmospheric specific intensity,
for which the brightness temperature is given by
Eq.~\ref{eq:Iatm2Tatm}.
and is tabulated along with $\tau$ in Table~\ref{tab:seasonsDependence}.
An obvious consequence of considering a multi-layer atmosphere is that
it does not have a single physical temperature,
because it is a combination of multiple
components, each having a different temperature
(Eq.~\ref{eq:Iatm}).
A multi-layer atmosphere however can be assigned
a single-layer atmosphere equivalent temperature, $\TSL$, that yields:
\begin{equation}
\label{eq:TSL}
\Iatm= B_\nu(\Tcmb) e^{-\tau(z_d)} + B_\nu(\TSL(\nu)) \bigl( 1-e^{-\tau(z_d)}\bigr).
\end{equation}
Combining Eq.~\ref{eq:Iatm} and Eq.~\ref{eq:TSL} gives
\begin{eqnarray}
\label{eq:TSL3}
\TSL(\nu) &=& \frac{h \nu}{k_B \ln{D}},\\\nonumber\\
D&=&  \frac{2 h \nu^3}{c^2} \frac{1-e^{-\tau(z_d)}}{\Iatm - B_\nu(\Tcmb)e^{-\tau(z_d)}}+1.\nonumber
\end{eqnarray}
For a single layer atmosphere $N=1$
and $\TSL=T_1$, and it is frequency independent. In general, however,
$\TSL$ depends on frequency because in different 
layers the temperature $T_i$ (Eq.~\ref{eq:Iatm}) is weighted by coefficients
$\sim (1-e^{-\tau_i})$ that depend on frequency
differently than $(1-e^{-\tau})$ (Eq.~\ref{eq:radiativeTransferLTE}),
and hence the frequency dependence does not cancel.
$\TSL$ is obviously season dependent too, and 
it can be readily derived from  Eq.~\ref{eq:TSL3} using
Eq.~\ref{eq:Iatm2Tatm}, $\Tatm$ and
$\tau$ estimates from Table~\ref{tab:seasonsDependence}. However, in
the optically thin limit, $\TSL$ is a sensitive function of
$\Tatm$ and $\tau$.  $\TSL$ is also a weak function of zenith distance. At
zenith, in the optically thin limit, it is a combination of all
atmospheric layers, but near the horizon the lowermost atmospheric
layers begin to dominate all other contributions.  Within the
flat-atmosphere model, $\TSL$ will reach the ground-level
atmospheric temperature at the horizon, but it will still vary from
one season to another.  In the optically thick limit,
$\TSL=\Tatm$ and it also loses its frequency dependence because in
this limit, the geometry is reduced to a single layer case.

\begin{figure}
\centering
\includegraphics[width=0.48\textwidth]{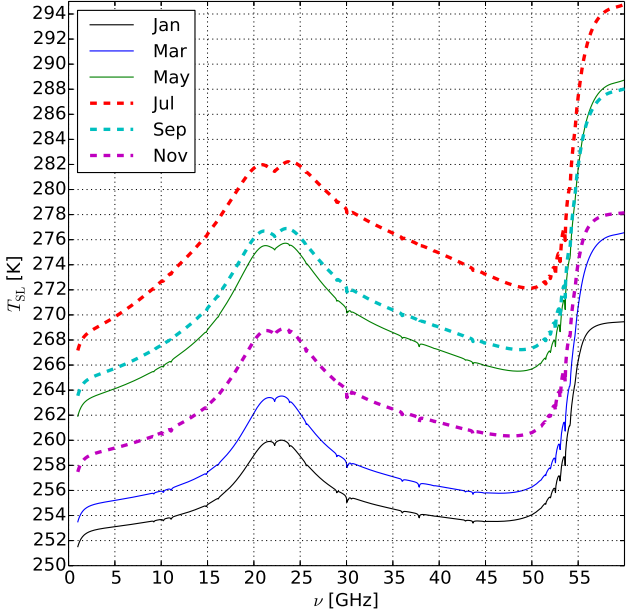}
\caption{Single-layer atmosphere equivalent temperature resulting from
the best fit model obtained from the combined PWV data samples
(Table~\ref{tab:PWVchisq}). The thin (solid) lines from the bottom to the top 
correspond to January, March and May, and the
thick (dashed) lines from the top to the bottom are calculated for July, September and November.}
\label{fig:TSL}
\end{figure}

In Fig.~\ref{fig:TSL} we plot the calculated $\TSL$ temperatures.
Individual curves result from the best
fit atmospheric models calculated for the selected months.
The spectra were calculated
using Eq.~\ref{eq:TSL3}, but due to the strong sensitivity of $\TSL$
to $\tau$ and $\Tatm$ the values cannot be constrained
to a precision better than O(10) K from the
current data.\footnote{For example, a variation 
of $\tau$ by $10^{-3}$ at 30 GHz in July implies a $\sim 4$ K change in
$\TSL$ (Table~\ref{tab:seasonsDependence}).} The dents coincide with the resonance frequencies of water
and ozone lines.

\begin{figure}
\centering
\includegraphics[width=0.48\textwidth]{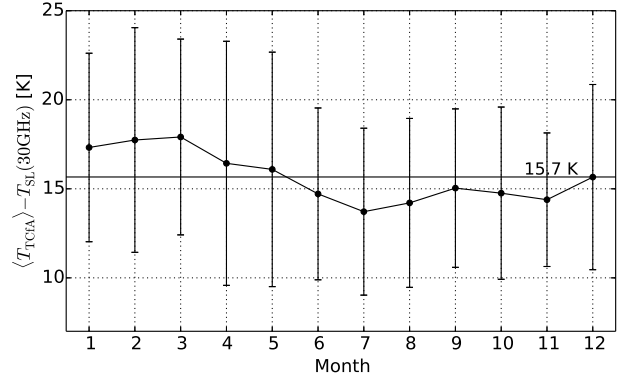}
\caption{Relationship between the single-layer atmosphere equivalent
temperature, $\TSL$, at 30 GHz, at the zenith (see Eq.~\ref{eq:TSL3}) and
the ground-level month-median atmospheric temperature at TCfA.  The
difference between the two is consistent, within $\pm 1\sigma$ error
bars, with a constant offset of about 15.7 K.  The error bars
represent the monthly ground-level temperature dispersion at TCfA.
The median and standard deviation of the ground-level atmospheric
temperatures are calculated using the meteorological data introduced
in Sec.~\ref{sec:meteoData}. The error bars are large as they
include diurnal temperature variations.}
\label{fig:TSL30GHz}
\end{figure}

For any given frequency the single-layer atmosphere equivalent
temperature ($\TSL$) should be most affected by the physical
temperatures of layers that are most responsible for absorption of
radiation at that frequency.  About 50\% of total WV contribution to
$\Tatm$ takes place above altitudes 1-2 km (where the temperature is
at least 5-10 K below the ground temperatures). At these altitudes
there is also the biggest concentration of WV. Similarly, the typical
height scale for $\sim 50$\% of oxygen absorption is about $\sim 5$ km
with typical temperatures $\sim 30$ K lower than ground
temperatures.\footnote{Rough estimates found using the best-fit model
in July.}  Since at 30 GHz the contributions to $\Tatm$ from oxygen
and WV are similar, the expected single-layer atmosphere equivalent
temperature should be about $\sim 20$ K lower than the ground
temperature. The result of a numerical calculation at 30 GHz is shown in
Fig.~\ref{fig:TSL30GHz}.  Clearly, for the location of TCfA a
reasonable estimate of $\TSL$, at 30 GHz, can be found from the ground
level temperatures, offset by $\sim 15.7$ K.  The offset for other
frequencies can be inferred from Fig.~\ref{fig:TSL}. Given the
sensitivity of $\TSL$ (Eq.~\ref{eq:TSL3}) to $\tau$, the offset
ground-level atmospheric temperature may provide a better constraint
than measurements of $\tau$ and $\Tatm$.

As mentioned earlier, $\TSL$ also depends on zenith distance, but we
find that the dependence is rather weak.
Up to $z_d=75^\circ$, $\TSL(z_d)$ is roughly constant
and fixed at its zenith value, with accuracy better than
$\{0.2,0.3,1.5,0.8\}$ K at frequencies $\nu=\{5,15,22,30\}$ GHz in July.
Deviations from the zenith values are even smaller in January.

\subsubsection{Planar atmosphere}
\label{sec:planar}

Radio telescopes with an Az-El mount cannot efficiently observe sources
towards the zenith, thus from practical reasons the dependence of $\Tatm$
on zenith distance ($z_d$) is important.  Based on the Bemporad's
air-mass--zenith-distance fitting formula
(see \cite{Schoenberg1929} or \cite{Wilson2009}), it can be seen that the
assumption of flatness of the atmosphere -- i.e. the atmosphere being
composed of flat layers stacked one over another -- is consistent with
$\sec(z_d)$ scaling to within 0.25\% (3\%) up to $z_d<60^\circ (80^\circ)$.
Within this approximation (which we use) the atmospheric optical depth
is given by
\begin{equation}
\tau(z_d) = \tau(0) \sec(z_d),
\label{eq:tau-zd-relation}
\end{equation}
where $\tau(0) \equiv \tau(z_d=0^\circ) $.
We assess the accuracy of this approximation by means of
radiative transfer through atmospheric
layers, whose thicknesses are increased 
to match ray path lengths travelling through spherical layers
at given zenith angle.
As before, we use the same setup
of $N=300$, layers distributed as discussed in
Sec.~\ref{sec:transfer}, but in this case each layer is
assumed to be located between geocentric
radii $R_i$ and $R_j$ and hence its thickness $h_{ij}=R_j-R_i$
at the zenith angle $z_d$ (measured from the surface of the Earth),
is given by:
\begin{eqnarray}
\label{eq:nonPlanar}
h_{ij}(z_d)&=&\left( R_i^2 + R_j^2 - 2 (R_\oplus^2+C_i C_j) \sin^2(z_d) \right)^{1/2}\\
\nonumber\\
C_i&=&\sqrt{R_i^2 \csc^2(z_d)-R_\oplus^2}\nonumber\\
\nonumber\\
C_j&=&\sqrt{R_j^2 \csc^2(z_d)-R_\oplus^2},\nonumber
\end{eqnarray}
where we assumed spherical Earth.

\begin{figure}
\centering
\includegraphics[width=0.48\textwidth]{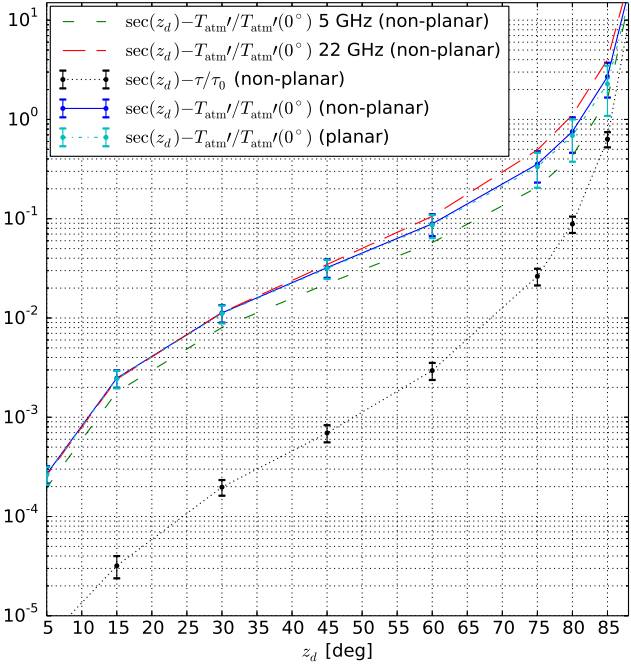}
\caption{Mean zenith distance dependence of $\tau/\tau(0^\circ)$ (black,dotted), and
$\Tatm'(z_d)/\Tatm'(0^\circ)$ (blue,solid and cyan,dash-dotted) for the best-fit local atmospheric
model. 
Error bars represent $\pm 1\sigma$ deviation from the mean
due to frequency and monthly dependence.
The short (long) dashed lines trace the season averaged dependence at
5 GHz (22 GHz). }
\label{fig:TatmTatm0}
\end{figure}

For each month, we use the best-fit atmospheric profile and calculate
$\tau(z_d)$ relations for a range of zenith distances, at the
frequencies $\nu [{\rm GHz}]=\{5,15,22,30\}$.  Then, we normalise
$\tau(z_d)$ at the zenith and average between different frequencies and
seasons.
The differences between individual
months (January and July) and frequencies vary between 40 and 53
at the horizon, and between 30 and 35
at $z_d=89^\circ$, and between 10.6 and 11
at $z_d=85^\circ$.

The following fourth order in $\sec(z_d)$ fitting function provides a
good fit to the reconstructed average relation, with the maximal
deviation below $0.3\%$ at $z_d=89^\circ$:
\begin{eqnarray}
\label{eq:airMassFit}
\left\langle\frac{\tau(z_d)}{\tau(0)}\right\rangle=1.001\sec(z_d)
- 3.739\cdot 10 ^{-4} \sec^2(z_d)\\
\nonumber\\
- 4.666\cdot 10 ^{-4} \sec^3(z_d) +6.0366\cdot 10 ^{-6}\sec^4(z_d)\nonumber.
\end{eqnarray}
We do not use $z_d=90^\circ$ data point to avoid infinite values,
but since the effects of refraction are not taken into
account, the usability of the formula is effectively reduced down to $z_d\lesssim
85^\circ$, where the dispersion due to a seasonal and frequency dependence
becomes similar to uncertainties due to neglected refraction.

According to the fitting formula, the assumption of the flat atmosphere
(Eq.~\ref{eq:tau-zd-relation}) is accurate to within 0.16
(1.6)\% at $z_d=60^\circ$ ($80^\circ$), where
$\left\langle \tau(z_d)/\tau(0)\right\rangle=1.997\pm
0.001$ ($5.670\pm 0.025$) and where the error bars represent the
$1\sigma$ dispersion due to season and frequency dependence.

To address the problem described at the beginning of this section, we
calculate the average
$\Tatm'(z_d)/\Tatm'(0)$ relations for the planar and non-planar
atmospheres (Fig.~\ref{fig:TatmTatm0}).  This is useful because radio
telescopes are directly sensitive to increases of antenna temperature.
The relations are derived according to the best fit atmospheric model
(last column of Table~\ref{tab:PWVchisq}).  The $\tau/\tau(0)$
relation for the planar case is given by $\sec(z_d)$ and is not
plotted, whereas the average $\tau/\tau(0)$ for the non-planar case
is given by Eq.~\ref{eq:airMassFit} (black/dotted line in Fig.~\ref{fig:TatmTatm0}).
The zenith distance dependence of $\Tatm'(z_d)/\Tatm'(0)$
is given by the following fitting formula,
which approximates the month and frequency averaged relation
below $z_d=89^\circ$ (dash-dotted/cyan line in Fig.~\ref{fig:TatmTatm0}): 
\begin{eqnarray}
\label{eq:TatmZDfit}
\left\langle\frac{\Tatm'(z_d)}{\Tatm'(0)}\right\rangle=1.022\,\sec(z_d)
- 3.203\cdot 10 ^{-2}\,\sec^2(z_d)\\
\nonumber\\
+ 9.625\cdot 10 ^{-4} \sec^3(z_d) -1.059\cdot 10 ^{-5}\sec^4(z_d)\nonumber,
\end{eqnarray}
where $\Tatm'$ is atmospheric brightness temperature calculated for the case when the
CMB is not present as a source term in the radiative transfer
equation, and $\Tatm'(0)\equiv\Tatm'(z_d=0^\circ)$.

Clearly, the 
difference of the mean $\Tatm'(z_d)/\Tatm'(0)$ from $\sec(z_d)$ scaling is
$\sim 0.1\,(0.4)$ or alternatively:
$(\sec(z_d)-\Tatm'/\Tatm'(0))/\sec(z_d)\approx 4\%\, (9\%)$ at
$z_d=60^\circ\, (75^\circ)$, where the averaging is done over twelve months
and the four considered frequencies (Fig.~\ref{fig:TatmTatm0}).
Looking into individual frequencies, the month averaged
$\sec(z_d)-\Tatm'/\Tatm'(0)\approx\{0.06,0.09,0.11,0.10\}$ for
$z_d=60^\circ$ at $\nu=\{5,15,22,30\}$ GHz.  The effects due to
planarity are more than order of magnitude smaller.  The difference
between the averaged $\tau/\tau(0)$ and $\Tatm'/\Tatm'(0)$
scalings increases as the optical depth escapes beyond the optically
thin limit. This is discussed in the next section.

\subsubsection{Optically thin atmosphere}
\label{sec:thin}
Within the RJ approximation $B_\nu(T)=2\nu^2 k_B T/c^2$, and
Eq.~\ref{eq:Iatm} becomes:
\begin{equation}
\Tatm(z_d) = \Tcmb e^{-\tau(z_d)} + \sum_{i=1}^N T_i \bigl( 1-e^{-\tau_i(z_d)}\bigr).
\label{eq:TatmRJ}
\end{equation}
The approximation of optically thin atmosphere is often exploited
by utilising the first order expansion in $\tau$ 
\begin{equation}
\Tatm^{(1)}(z_d) = \Tcmb (1- \tau(z_d)) + \TSL \tau(z_d),
\label{eq:Tatm-zd-approx2}
\end{equation}
where, as before, we utilised the single layer equivalent temperature
at the zenith (see below).  We checked that in order to visualise the
shortcomings of using the optically thin atmosphere approximation the
differences due to introducing the RJ approximation are unimportant.

\begin{figure}
\centering

\includegraphics[width=0.48\textwidth]{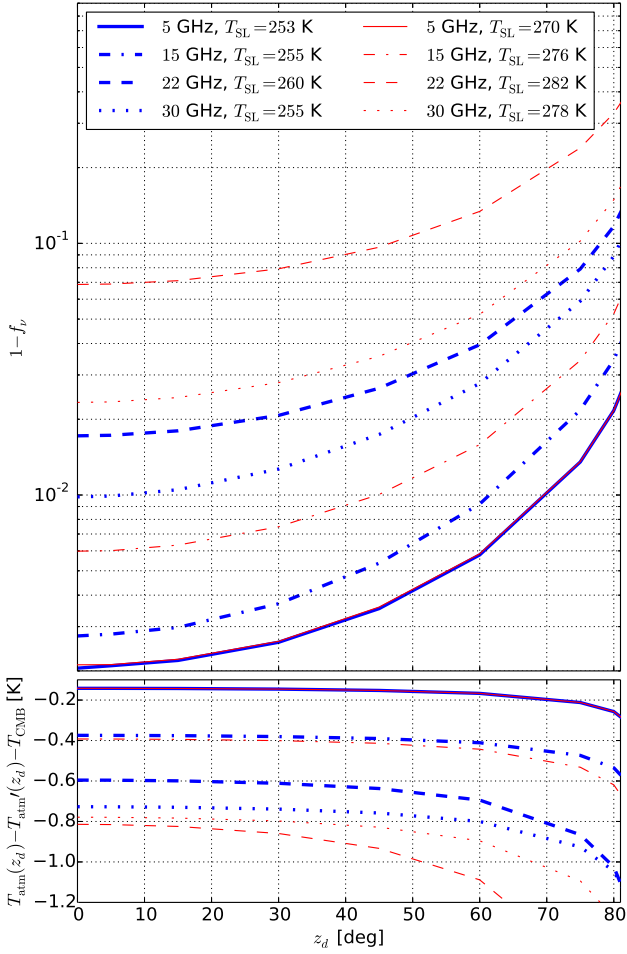}
\caption{({\it Top panel}) 
Accuracy of the atmospheric brightness temperature linear order
expansion (Eq.~\ref{eq:TatmTatmlin}) in January (blue/thick) and July
(red/thin).  The best-fit model ($\pH=0.62$) from the combined PWV data
samples (Table~\ref{tab:PWVchisq}) is used.  For each frequency the
zenith $\TSL$ values are provided in the plot legend.
({\it Bottom panel}) Accuracy of $\Tatm\approx \Tatm\prime + \Tcmb$
approximation expressed as a difference between the actual $\Tatm$
and the sum of constituents considered separately: $\Tcmb$ and
atmospheric brightness temperature derived without the CMB as a
source term ($\Tatm'$).
}
\label{fig:Tatm-zd}
\end{figure}

Depending on what is being measured, $\Tcmb$ contribution is sometimes
treated independently from $\Tatm$, as if the CMB was not processed by
the altitude dependent absorption.  Below, we quantify this
approximation, but in general, since the CMB propagates through an
absorbing and emitting medium, the two contributions (one from the
atmosphere and the other from the CMB) cannot be treated separately. For
this reason, our constraints on $\Tatm$ include the CMB as a source
term in Eq.~\ref{eq:radiativeTransfer2}.
Neglecting the CMB as a source term,
Eq.~\ref{eq:Tatm-zd-approx2} for the case of
flat atmosphere (Eq.~\ref{eq:tau-zd-relation}) 
becomes:
\begin{equation}
\Tatm^{(1)}(z_d)= \TSL \tau(0)\sec(z_d),
\label{eq:Tatm-zd-approx3}
\end{equation}
and what follows is that for a single layer atmosphere
\begin{equation}
\Tatm'(60^\circ)= 2\, \Tatm'(0^\circ),
\label{eq:Tatm-airmass-assumption}
\end{equation}
where $\Tatm'(z_d)$ is the atmospheric brightness temperature for $\Tcmb=0$ K.
This approximation is often used in tipping scan measurements.

Increasing the air mass, by looking at greater $z_d$ angles,
the linear-order approximation in Eq.~\ref{eq:Tatm-zd-approx3} 
must fail at high values
of $\tau$ as $\Tatm$ (or $\Tatm'$) cannot reach infinity.
The ratio of
Eq.~\ref{eq:TatmRJ} and Eq.~\ref{eq:Tatm-zd-approx2}:
\begin{equation}
f_\nu(\Tcmb,\TSL,\tau_0,z_d) = \Tatm/\Tatm^{(1)}
\label{eq:TatmTatmlin}
\end{equation}
gives a factor by which $\Tatm$ is underestimated
by using the linear-order single-layer approximation.
Notice that Eq.~\ref{eq:Tatm-zd-approx2}, and consequently Eq.~\ref{eq:Tatm-zd-approx3},
introduce yet another,
commonly used approximation, that $\TSL$ does not depend on $z_d$, but
instead is fixed at its zenith value.
As discussed in Sec.~\ref{sec:SL}, this is quite a reasonable approximation
for a range of $z_d$s.
$\Tatm$ in Eq.~\ref{eq:TatmTatmlin} is calculated numerically
within the flat atmosphere approximation, and $\Tatm^{(1)}$ is calculated
according to Eq.~\ref{eq:Tatm-zd-approx2} with $\tau(z_d)$ derived
numerically and with $\TSL$ calculated from Eq.~\ref{eq:TSL3}.

The factor $f_\nu$ depends on $\tau$,
frequency, and vertical temperature profile of the atmosphere.
Assuming a flat and single-layer atmosphere without the CMB, it is
easy to see that the factor becomes independent of atmospheric temperature
and at $z_d=60^\circ$ amounts to
$f(\Tcmb=0\, {\rm K},\tau_0)=(1-e^{-\tau_0 \sec(z_d)})/(\tau_0\sec(z_d))
\approx 0.965 (0.94)$ at 30 GHz for the best-fit value of
$\tau_0=0.036 (0.061)$ in January (July) (Table~\ref{tab:seasonsDependence}).

In the bottom panel of Fig.~\ref{fig:Tatm-zd} errors of considering
$\Tcmb$ independently from $\Tatm$ are quantified. These are the lowest at
low frequencies, small optical depths and small zenith angles (see plot description
for details).

\subsubsection{Relation to observations}
\label{sec:relobs}
The approximations discussed in the previous sections may
affect the radio source flux density observations at the levels of
few to several per-cent.

In particular, estimates of the atmospheric absorption corrections may
be biased depending on the assumed approximations.  Flux-density
absorption corrections within the flat and optically thin
atmosphere model are given by:
\begin{equation}
S_t = S_m e^{\tau_0 \sec(z_d)}\approx S_m \bigl(1+\tau_0 \sec(z_d)\bigr)
\label{eq:absCorr}
\end{equation}
where $S_t$ is the true flux density and $S_m$ is the measured flux
density.  These corrections require an estimate of
$\tau=\tau(\nu,z_d,m)$, which is a function of frequency ($\nu$), zenith
distance ($z_d$), and month ($m$).

With a single-dish radiometers $\tau$ is typically estimated by measuring
the system temperature components at the zenith and at $z_d=60^\circ$ with
an implicit assumption of the validity of
Eq.~\ref{eq:Tatm-airmass-assumption}.  In the simplest case, when
detector linearity is assumed and the spillover, side-lobe and
ground pick-up contributions are neglected, in the RJ approximation,
the measurement can be defined by the following system of linear
equations:
\begin{equation}
\begin{array}{lll}
c_f V(0^\circ) & = &     \Tatm'(0^\circ) + T_{\rm{rx}} +\Tcmb + \Delta T_\nu(0^\circ) \\
c_f V(60^\circ) & = &  A_m(60^\circ) \Tatm'(0^\circ) + T_{\rm{rx}} +\Tcmb + \Delta T_\nu(60^\circ)\\
c_f V_{\rm abs} & = &        T_{\rm{abs}} + T_{\rm{rx}}  , 
\end{array}
\label{eq:TatmMeasure}
\end{equation}
where $T_{\rm rx}$ is the receiver noise temperature, $T_{\rm abs}$ is
the absorber temperature and $c_f$ is a conversion factor from the
voltage, measured at the square-law detector, to the units of antenna
temperature, and $V(z_d)$ is the measured voltage at $z_d$.
$\Delta T_\nu$ is a correction factor weakly dependent on the zenith distance
(Fig.\ref{fig:Tatm-zd} bottom panel).
Usually, at $z_d=60^\circ$ the $A_m=2$ (and $\Delta T_\nu=0$)
is assumed, in accordance with the flat atmosphere expectation.

Substituting thus derived value of $\Tatm'$ into
Eq.~\ref{eq:Tatm-zd-approx3} with an assumed value of $\TSL$ gives an
estimate of $\tau$. This is a possible source of systematical effects.
First, as discussed in Sec.~\ref{sec:SL}, $\TSL$ follows ground level
temperatures and fixing its value will lead to season dependent
systematical effects.
Secondly, Fig.~\ref{fig:Tatm-zd} indicates that
using an approximation given by Eq.~\ref{eq:Tatm-zd-approx3} leads to a slight,
but systematical underestimation of $\tau$ and $S_t$. In practice,
absorption corrections are applied to both, the target source at
$z_d^{\rm src}$ and the chosen flux calibrator at the zenith distance
$z_d^{\rm cal}$, hence the effects of the approximation in
Eq.~\ref{eq:Tatm-zd-approx3} on the relative flux density will
cancel, as long as the calibrator and the target source happen to be
observed at the same $z_d$.  This is however almost never the case,
and so the effects of the approximation will propagate onto the flux
density estimates through factors $\sim (1+\tau_0 \sec(z_d^{\rm
src}))/(1+\tau_0 \sec(z_d^{\rm cal}))$. Since the calibrator and
target sources are observed at different elevations, this effectively
leads to an increased variance of flux density estimates for
individual sources, disparities between sources of the same intrinsic
flux density, and possibly to systematical effects depending on
source declination with respect to the calibrator.

\subsection{$\Tatm/\tau$-PWV scaling relation}
\label{sec:TatmPWV}

\begin{figure}
\centering
\includegraphics[width=0.48\textwidth]{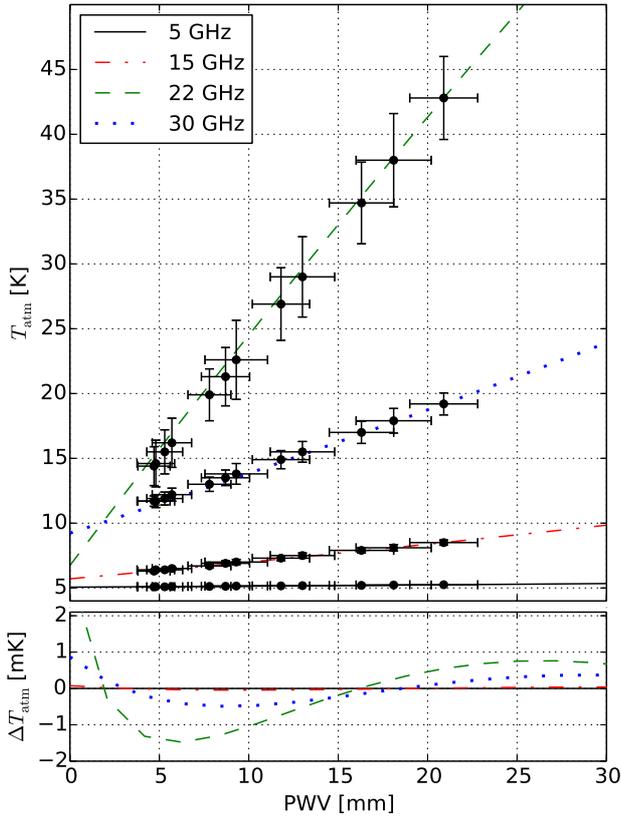}
\caption{({\it top panel}) $\Tatm$-PWV scaling relations given by
Eq.~\ref{eq:TatmPWV} and Table~\ref{tab:TatmPWV} (lines).  The four
sets of points over each of the lines represent the best-fit model
brightness temperatures from Table~\ref{tab:seasonsDependence}.
({\it bottom panel}) Residues between the calculated scaling
relations described in Sec.~\ref{sec:TatmPWV}, and their respective
fitting functions. }
\label{fig:TatmPWV}
\end{figure}

In clear sky conditions and for a given vertical profile of PWV,
there is a 1:1 relation between the measured PWV and $\Tatm$.  In this
section we derive $\Tatm$-PWV and $\tau$-PWV scaling relations
for January and July, using
the best fit model vertical profile estimates from the combined data samples
(Table~\ref{tab:PWVchisq}) . In each profile
the PWV content is scaled at all altitudes by constant factors, and
the corresponding $\Tatm$ and $\tau$ are recorded.  
We calculate the scaling relations at frequencies 5,15,22 and 30 GHz and
find that the following third order polynomial provides a good fit
with negligible residues:
\begin{equation}
X = \sum_{i=0}^3 x_i \biggl(\frac{{w}}{1\, {\rm mm}}\biggr)^i
\label{eq:TatmPWV}
\end{equation}
where $X=\{\Tatm, \tau\}$ and $x_i=\{a_i,b_i\}$.  The
$a_i$ and $b_i$ coefficients are summarised in Table~\ref{tab:TatmPWV}.
Terms beyond the linear order describe the subtle effects of pressure
line broadening and self-induced water vapour continuum absorption,
both non-linearly dependent on ${\rm H_2O}$ VMR
\citep{Paine2015communication1}.  For $\tau$ the scaling is quadratic
and so we use the second order polynomial.
In Fig.~\ref{fig:TatmPWV} we visualise $\Tatm$
\footnote{According to our notation $\Tatm$
includes the CMB as a source term (Sec.~\ref{sec:transfer}).}
scaling
relations and the residual errors
between these relations and their fitting formulas
(Eq.~\ref{eq:TatmPWV}).

\begin{table*}
\caption{\label{tab:TatmPWV}
$\Tatm$-PWV and $\tau$-PWV fitting formula coefficients
(Eq.~\ref{eq:TatmPWV}).  The numbers are written in scientific
notation with the decimal exponent in parentheses.  The 'error'
column contains standard deviations of the difference between the
derived scaling and the corresponding fitting formula.  }
\Beginruledtabular
\begin{tabular}{rcccccccc}
\BeginruledtabularSec
&\multicolumn{8}{c}{January} \\
$\nu$ [GHz]  & $a_0 \,{\rm [K]}$   & $a_1 \, {\rm [K/mm]} $  & $a_2  \,{\rm [K/mm^2]}$   & $a_3  \, {\rm [K/mm^3]}$   & error [mK] 
& $b_0 $   & $b_1 \, {\rm [mm^{-1}]} $  & $b_2  \,{\rm [mm^{-2}]}$    \\
5 & $ 5.055 (+00) $ & $ 8.104 (-03) $ & $ 7.997 (-05) $ & $ -2.841 (-09) $ & $ 2.441 (-06) $ & $ 9.350 (-03) $ & $ 3.108 (-05) $ & $ 3.047 (-07) $  \\
15 & $ 5.738 (+00) $ & $ 1.282 (-01) $ & $ 6.849 (-04) $ & $ -3.392 (-07) $ & $ 2.551 (-04) $ & $ 1.209 (-02) $ & $ 4.921 (-04) $ & $ 2.741 (-06) $  \\
22 & $ 6.833 (+00) $ & $ 1.644 (+00) $ & $ -3.827 (-03) $ & $ 8.033 (-06) $ & $ 1.434 (-01) $ & $ 1.650 (-02) $ & $ 6.391 (-03) $ & $ 5.896 (-06) $  \\
30 & $ 9.477 (+00) $ & $ 4.603 (-01) $ & $ 2.396 (-03) $ & $ -4.450 (-06) $ & $ 6.191 (-03) $ & $ 2.729 (-02) $ & $ 1.792 (-03) $ & $ 1.096 (-05) $  \\

&\multicolumn{8}{c}{July} \\
5 & $ 5.062 (+00) $ & $ 7.640 (-03) $ & $ 5.368 (-05) $ & $ -2.070 (-09) $ & $ 3.662 (-04) $ & $ 8.820 (-03) $ & $ 2.728 (-05) $ & $ 1.901 (-07) $  \\
15 & $ 5.705 (+00) $ & $ 1.245 (-01) $ & $ 4.551 (-04) $ & $ -2.370 (-07) $ & $ 3.417 (-02) $ & $ 1.125 (-02) $ & $ 4.450 (-04) $ & $ 1.710 (-06) $  \\
22 & $ 6.739 (+00) $ & $ 1.825 (+00) $ & $ -4.879 (-03) $ & $ 6.163 (-06) $ & $ 9.216 (-01) $ & $ 1.515 (-02) $ & $ 6.616 (-03) $ & $ 3.679 (-06) $  \\
30 & $ 9.230 (+00) $ & $ 4.455 (-01) $ & $ 1.555 (-03) $ & $ -3.130 (-06) $ & $ 3.457 (-01) $ & $ 2.472 (-02) $ & $ 1.612 (-03) $ & $ 6.840 (-06) $  \\

\EndruledtabularSec
\end{tabular}

\Endruledtabular
\end{table*}

The fitting formulas defined in Eq.~\ref{eq:TatmPWV} were derived
using the best-fit models for January and July by scaling the WV
content, hence the $\Tatm$ (or $\tau$) predicted by these formulas
will not exactly match the best-fit model values from
Table~\ref{tab:seasonsDependence}.  This is because
the vertical structure of the atmosphere is month dependent. In order
to visualise the amplitude of these differences, for each frequency,
we over-plot the twelve $\Tatm$ values form
Table~\ref{tab:seasonsDependence} atop the fitting relation obtained
for July (Fig.~\ref{fig:TatmPWV}).  $\Tatm$ from different months are consistent within
$1\sigma$ error bars quoted in Table~\ref{tab:seasonsDependence}, but a
better consistency is reached at high PWV levels, typical for summer
months, as expected.  The year-average PWV-$\Tatm$ scalings, which result
from fitting the data points from Fig.~\ref{fig:TatmPWV}, are
given by $A$ and $B$ coefficients in Table~\ref{tab:seasonsDependence}.

$\Tatm$-PWV scaling relation can be used to estimate 
$\Tatm$ instabilities due to time varying column PWV. The amplitude of
temporal fluctuations of the column PWV,
under the frozen turbulence hypothesis,
should constrain the spatial power spectrum of
water vapour distribution at the scales $L$, roughly corresponding to
$v_w \Delta t$, where $\Delta t$ is the time interval between
subsequent PWV measurements,
and $v_w$ is the wind speed at relevant altitudes.
Alternatively, on the same
assumptions, two instantaneous measurements of PWV, performed at two
locations separated by distance $L$, probe the turbulent spectrum of
PWV at the scales roughly corresponding to the spatial separation of
the observation sites.  While the spectrum of the atmospheric PWV
fluctuations is beyond the scope of this paper, we discuss a rough
estimate of the amplitude of the $\Tatm$ variations in clear sky
conditions, based on the sun photometer measurements of
the column PWV.

Since the $\Tatm$-PWV scaling relation is dominated by the linear
terms ($a_0$ and $a_1$) the $a_1$ term can be thought of as an
approximation of the derivative $\frac{ \d \Tatm}{\d w}$ which
quantifies $\Tatm$ response to variations in PWV
($w$). Table~\ref{tab:TatmPWV} shows that PWV variation of $\sim 0.2$
mm ($\sim 0.3$ mm) would cause $\Tatm$ variation of the order
$\Delta\Tatm \approx 0.09\, (0.14)$ K at 30 GHz in January (July).
These are the amplitudes of the PWV variations in clear sky
conditions we actually observe at $\sim 1$-hour time
scales.\footnote{The amplitude of the PWV variations quoted are
examples from a single day observations in January and July, but
the amplitude depends on the considered time-scale,
and also somewhat varies from one day to another. }
Clearly, low column
PWV values also result in a more stable atmospheric brightness
temperature.  By the Kolmogorov's atmospheric turbulence model, the
brightness fluctuations are characterised by a steep spectrum over a
wide range of spatial frequencies, and the overall variance is
dominated by the largest scales. For the wind speed of the order $O(1)$ m/s
typical for a calm sky without frontal activities, 1-hour
time scale corresponds to $L\sim O(10)$ km length scales, which coincide
with the largest
scales from the inertial sub-range, in which the atmospheric turbulence driven
cascade of kinetic energy transport takes place.  Incidentally, this
rough estimate remains in agreement with theoretical predictions from
the atmospheric turbulence model \citep{Baars2007}, which for the same
frequency at the sea level gives $\Delta\Tatm\approx 0.0055\,
\Tatm$, that is $\Delta\Tatm \approx \{0.06,0.11\}$ K at 30 GHz for January and
July respectively (Table~\ref{tab:seasonsDependence}).
This is also
qualitatively compatible with an independent analysis involving the
near-ground RH variability and wind speed measurements (Lew 2016, in
preparation).

\subsection{Accuracy limits}
\label{sec:accuracy}

The precision to constrain the local, mean $\Tatm$ and $\tau$
range from $\sim 0.2\%$ at 5 GHz to $\sim 13\%$ at 22 GHz (see
Table~\ref{tab:seasonsDependence} for January).  These uncertainties
refer to the monthly mean estimates, rather than to
individual PWV measurements, although the dispersion of the individual
measurements affects the $\chi^2$ value when model fitting.
A random PWV measurement may deviate significantly from
the mean even in clear sky conditions (e.g. see the scatter
from individual PWV measurements in Fig.~\ref{fig:PWV} or
Table~\ref{tab:PWV}).
In order to assess the accuracy with which 
the actual $\Tatm$ and $\tau$ can be constrained from the
statistical analysis of the climatology data,
we calculate the impact of
the greatest variations, observed in our PWV
measurements, on the values of $\Tatm$ and $\tau$.

The greatest observed variation of PWV in our sample occurs in June
(Table~\ref{tab:PWV}) and ranges from $w \sim 9$ mm to $w \sim 23$ mm
and the $1 \sigma$ variation is approximately $\pm 6.1$ mm.  Assuming
linearity of the PWV--$\Tatm$ relation ($a_1$ coefficients for July in
Table~\ref{tab:TatmPWV}), this translates to $1 \sigma$ temperature
variations $\Delta\Tatm \lesssim \pm \{0.05, 0.8, 11, 2.7\}$ K at $\nu
= \{5,15,22,30\}$ GHz, or by about
$\Delta\Tatm/\Tatm \approx \pm \{1, 9, 26, 14\}$ \% (see
Table~\ref{tab:seasonsDependence} for July). Clearly, low frequencies
are the least sensitive to WV variations, as expected.

\subsection{Clear sky detection}
\label{sec:clear_sky_detection}

\begin{figure}
\centering
\includegraphics[width=0.48\textwidth]{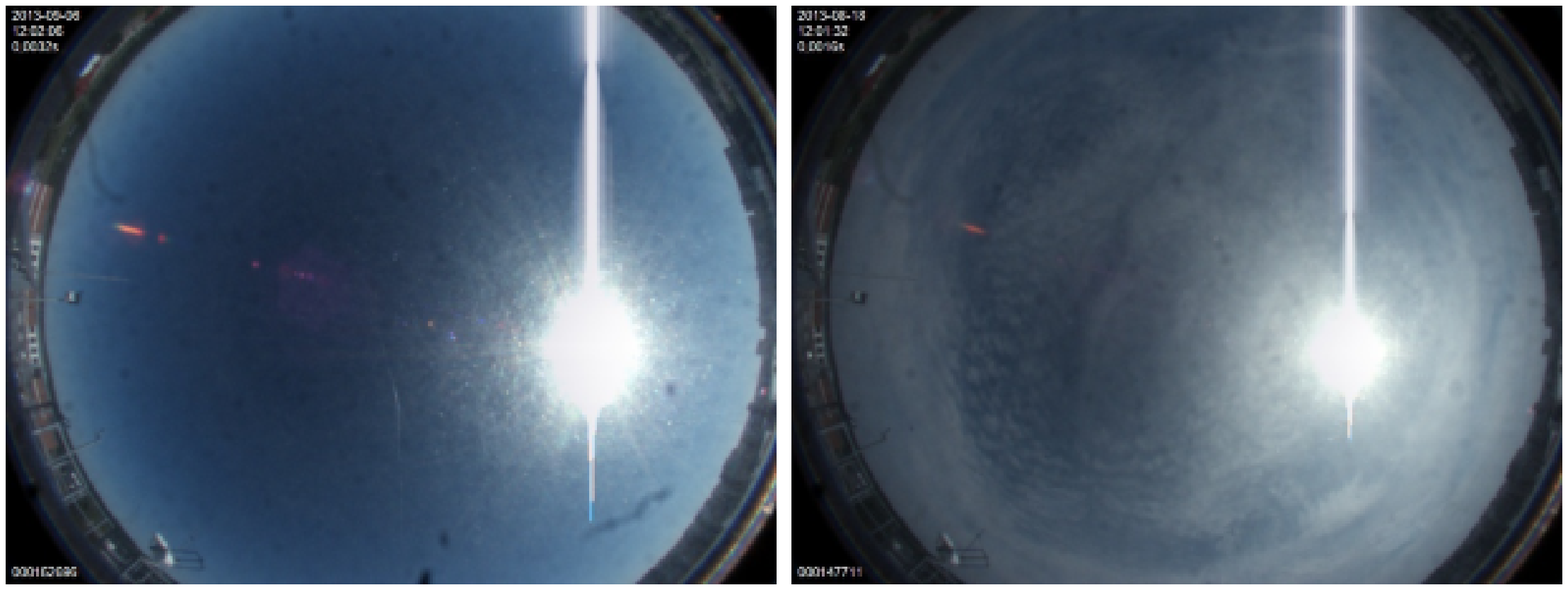}\\
\includegraphics[width=0.48\textwidth]{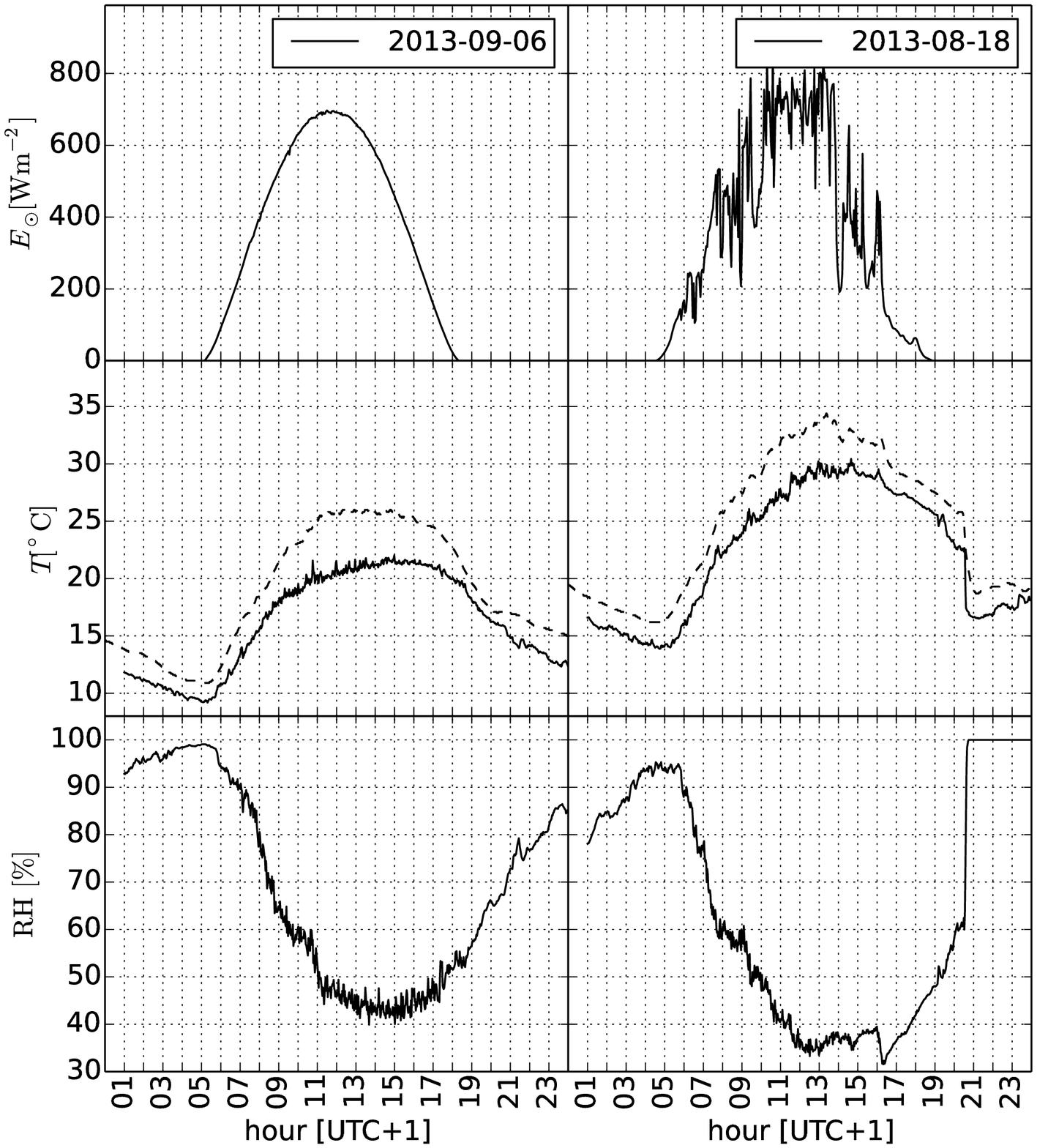}
\caption{Comparison of atmospheric parameters on two different days
identified as having clear sky conditions, selected on the basis of low RH values.
The left panels show
the correct selection (with an average 4-hour cloud cover
${\rm CI}= 0$), and the right panels show an example of a false-positive
selection (with an average 4-hour cloud cover ${\rm CI}\approx 6$).
From the top to the bottom the panels show the all-sky camera
picture around noon (UTC+1), the ground-level all-sky solar irradiance,
temperature and RH. The dashed lines trace the {\it CMP 22}
pyranometer sensor temperature. While the false-positive selection
occurred most likely due to the little impact of high clouds on
the near-ground RH, the impact of such a thin layer of high clouds
is clearly significant on the detected irradiance. }
\label{fig:clouds}
\end{figure}

By the analysis of the ground-level solar irradiance ($E_\odot$),
temperature ($T$) and all-sky images, we observe that in clear
sky conditions there is a good positive correlation between the solar
irradiance and air temperature.  At such times $E_\odot$ and $T$ are
smooth functions of time.  On the other hand, clouds passing through
the LOS towards the Sun generate a high frequency noise
(Fig.~\ref{fig:clouds} second row panels).  We anticipate that this feature
can be used for automatic cloud detection in
meteorological data analyses, but we defer details to a separate study.

In Sec.~\ref{sec:intro} we mentioned that for a fixed pressure level
the RH has a diurnal variation corresponding to temperature variations
of the atmospheric layer.  The RH variation anti-correlates with the
near-ground atmospheric temperature and also with the solar irradiance
(Fig.~\ref{fig:clouds}) detected at DMC.  However, IGRA data from Legionowo
(Sec.~\ref{sec:IGRA}) are recorded only twice a day (midday and
midnight) and it is impossible to analyse the stability of diurnal
variations of RH in an attempt to infer the cloud cover.  It is therefore,
interesting to see how feasible it is to select sub-samples of IGRA
data, that would statistically correspond to clear sky conditions,
given only few observational parameters: $T$,$P$ and $T_{\rm dew}$.
We will investigate this using local sky images archived at DMC and 
$T$,$P$ and RH measurements from TCfA, coarse averaged at $1$-hour time scale.
The distance between the two sites is $\sim 8$ km.

The amplitude of diurnal temperature
variations is the greatest in clear sky conditions.
This is because the radiative cooling of the
surface of the Earth is more efficient without heat-trapping
clouds.  Since the RH follows these variabilities in anti-phase, it
should be expected that at fixed temperatures a selection based on the
lowest daytime RH values should statistically correspond to clear sky
conditions.  On the other hand, a thick cloud cover tends to mitigate
the amplitude of day to night variations in temperature and humidity.
In this section we report results of a statistical analysis aimed to
verify the accuracy of this hypothesis.

The hourly averaged ground-level meteorological data (Sec.~\ref{sec:meteoData})
are screened by RH to form a sub-sample
of the 5\% driest conditions (sec.~\ref{sec:meteoData}).  In order to
mitigate effects of diurnal temperature variations we consider only
the samples obtained between hours 10:00 and 14:00 of the UTC+1 time. Such a
choice roughly corresponds to the times when the
temperatures should be the most stable (Fig.~\ref{fig:clouds}).
The selection typically picks out two
days per each month, depending on data completeness. Next, we visually
analyse the all-sky images for the selected
days and times and assign a mean cloud cover index for two cases:
(i) disregarding the distinction between high, medium and low clouds, and
(ii) ignoring high clouds i.e. treating them as clear sky.  The cloud
cover index (CI) can range from 0 for no clouds situation,
up to 8 for the full sky cloud cover. We disregard whether the cloud
cover is thin or thick, which is a conservative choice. The result is summarised in
Table~\ref{tab:clouds_detection}

\begin{table}
\caption{\label{tab:clouds_detection} Summary of the sensitivity of
the cloud detection algorithm by low ground-level RH values for the
two cloud cover categories: 'All' where all types of clouds are
considered, and ``Low and Medium'' for which high clouds
are treated as a clear sky. }
\Beginruledtabular
\begin{tabular}{cccccccc}
\BeginruledtabularSec
Cloud cover & &\multicolumn{6}{c}{True-Positive fraction [\%]} \\
sky fraction & Clouds\footnotemarkTAB{$^a$}: &\multicolumn{3}{c}{All} & \multicolumn{3}{c}{Low \& Med.} \\
(CI\footnotemarkTAB{$^c$})& Season\footnotemarkTAB{$^b$}: & C\&H & C & H & C\&H & C & H \\
$ \leq 0.125\, (1) $ & & 24 & 23 & 24 & 57 & 56 & 58 \\
$ \leq 0.250\, (2) $ & & 34 & 29 & 40 & 65 & 56 & 74 \\
$ \leq 0.375\, (3) $ & & 48 & 35 & 62 & 76 & 62 & 89 \\
\EndruledtabularSec
\end{tabular}
\Endruledtabular

\footnotetextTAB{$^a$Type of clouds considered. 'Low \& Med.' means that
all high clouds (if present) were treated as a clear sky.}
\footnotetextTAB{$^b$'H' - hot season: months from April to September, 'C' - cold
season: months from October to March, 'C\&H' - all year.}
\footnotetextTAB{$^c$CI - cloud cover index ranging from 0 to 8.
For example, $\rm{CI}=1$ corresponds to cloud cover sky fraction of $1/8=0.125$.}
\end{table}

Clearly, the selection by RH at the ground level has a high
false-positive (low true-positive) fraction when all types of
clouds are considered (see ``All'' column in
Table~\ref{tab:clouds_detection}).  However the situation significantly
improves when high clouds are treated as a clear sky.  This result is
easy to interpret.  There is generally a poor correlation between RHs
(or the corresponding ${\rm H_2O}$ VMRs) at different pressure levels
(altitudes) hence, there is no guarantee that selection by low
near-ground RHs will pick out days with no or little high cloud cover.
Even the straightest selection by RH (5\% driest days per month) does not
always select fully cloudless days.  This typically happens for
the months during which the sky is cloudy most of the time (such as
November in Poland).  An example of a false-positive detection is
shown in Fig.~\ref{fig:clouds}.  The two days were selected by the
algorithm as cloudless.  The RH variabilities (bottom panels) and their lowest
values are similar in both cases.  Yet, a visual inspection revealed
that the earlier day was contaminated by high clouds and classified as
${\rm CI}=6$ on the average, as opposed to the latter day classified
with ${\rm CI}=0$.  It seems however, that solar irradiance measured
directly to the Sun could be effectively used to detect clouds by
analysing the high frequency Fourier modes of the time domain signal.
\footnote{We measure the solar irradiance using the {\it CMP22}
pyranometer operating in the spectral range from 200 nm to 3600 nm
and within $\sim 180^\circ$ FOV.}  When treating high clouds as
clear sky, however, the selection of days with cloud cover below 0.375
is accurate in $\sim 76$\% of cases year-round and in $\sim 89$\% of
cases during the hot season (see Table~\ref{tab:clouds_detection}).  We
have not investigated the specificity of the estimator -- i.e.  we
have not estimated the fraction of clear sky days that eluded
detection, but we are aware of such cases.

On average, disregarding the cloud type, the fraction of days with
cloud cover $\leq 20$\% is $\sim 10\%$ year-round, $\sim 12$\% in the hot
season\footnote{See Table~\ref{tab:clouds_detection} for the
definitions of the hot and cold seasons.} and $\sim 8$\% in the cold season
for the RT32 region \citep{Wos2010}.  Based on the data from
Table~\ref{tab:clouds_detection} (interpolated for the cloud cover
$\leq 20$\%), in the case of all types of clouds, using a binomial
distribution with the number of trials fixed by the number of clear sky
day candidates selected by the algorithm, we estimate that the
probability of obtaining the reported true-positive rates
by chance is very low, $<10^{-4}$. Here, we assumed a
true-positive outcome as a successful trial, with the probability
defined by the aforementioned cloud cover statistics established by
the local climate.

While the algorithm of selecting clear sky conditions by low,
ground-level RHs yields a statistically significant result, we have
not investigated the true-positive fractions where only a single
measurement per day is available (case for IGRA data).  It is possible
however, that the algorithm can be improved by modifying the 
time range for data selection (here set between 10:00 and 14:00 of UTC+1), or by
altering the RH threshold (here set at 5\% of driest conditions), or
by including the long-term pressure variations. Due to the reasons
described in Sec.~\ref{sec:parametrization}, for the main analysis, we
employed external PWV measurements to match the data selection
criteria to the clear sky requirements.

\section{Discussion}
\label{sec:discussion}

We used PWV observations (Sec.~\ref{sec:PWVdata}) 
to match the mean PWV profile from the sounding data (Sec.~\ref{sec:WVprofiles})
to the expectations of clear
sky conditions. With the current parametrisation (Sec.~\ref{sec:parametrization}), however,
we are unable to relate the $\pH$
best fit value to some other observational quantity that is known
from meteorological statistics, hence the constrained value of the $\pH$
selection parameter is potentially useful only for the locations 
with similar climate.  Where local clear sky PWV measurements
are not available, a reasonable 
approximation of the clear-sky atmospheric WV profile could be found using the
nearest AERONET data (Sec.~\ref{sec:PWV}, Fig.~\ref{fig:PWV} and Fig.~\ref{fig:PWVchisq}).

Although filtering meteorological data by 
the lowest 5\% of the ground-level RH,
statistically, tends to pick up cloudless days 
(Sec.~\ref{sec:clear_sky_detection} and Table~\ref{tab:clouds_detection}),
such a single level selection does not correspond to the $0.05$ value
of the $\pH$ selection parameter since, by definition, $\pH$ filters 
the data coherently at all altitudes (Sec.~\ref{sec:parametrization}), therefore
such an association would strongly bias the statistics towards the driest conditions.

\section{Conclusions}
\label{sec:conclusions}

We use climatological data to reconstruct the vertical structure
of the atmosphere, constrain month-dependent profiles of precipitable water vapour
(PWV), and predict the atmospheric brightness temperature
($\Tatm$) and optical depth ($\tau$) at cm-wavelengths.  We
demonstrate that the nearly global coverage of the publicly available
climatological data enables investigation of almost every location world
wide, with the spatial resolution of a few hundred kilometres, on
average (Sec.~\ref{sec:data}).

We compare the month-dependence of the column PWV, reconstructed for
the location of the 32-metre radio telescope (RT32) located near Toru\'n
(Poland), with the AERONET data, collected at the closest station
located in Belsk. We find that the two are closely correlated
throughout the year, which supports the reliability of the PWV
reconstruction from sounding and ground-base meteorological data.

Bearing in mind the prerequisites of radio source continuum flux density
measurements at cm-wavelengths, we focus on radiative properties of
the atmosphere in clear sky conditions.  We present a
compilation of $\sim 17$ months of local PWV observations
collected in clear sky conditions 
using the MICROTOPS II sun photometer (Sec.~\ref{sec:PWV}).
We use these observations to devise
a selection criterion, which when applied to the climatological data,
enables us to reconstruct the vertical structure of the atmosphere that is
compatible with a cloudless sky.

Using the reconstructed clear sky PWV profile, and solutions of the
radiative transfer through the atmosphere, we constrain $\Tatm$ and
$\tau$ for the first time for the location of RT32
(Sec.~\ref{sec:Tatm}). We also establish PWV-$\Tatm$ and PWV-$\tau$
scaling relations (Sec.~\ref{sec:TatmPWV}) that can be used to
constrain atmospheric brightness temperature and optical depth in 
clear sky conditions, given an independent estimate of PWV (e.g. from
a local GPS station).  We estimate that in clear sky conditions,
the mean monthly values of $\Tatm$ and $\tau$, inferred from
climatology data, constrain the actual values to within $\pm \{1, 9,
26, 14\}$ \% (at $1\sigma$ CL) at $\nu = \{5,15,22,30\}$ GHz. These
estimates should also apply to other locations at similar latitudes and
a compatible climate (Sec.~\ref{sec:accuracy}).

We calculate the zenith distance ($z_d$) dependence of $\Tatm$ and
discuss $\lesssim 10\%$ effects regarding radio-source continuum flux
density measurement calibrations. We discuss the implications of using
optically thin, single-layer, and flat atmosphere approximations in
determining the optical depth and estimating corrections for
atmospheric absorption (Sec.~\ref{sec:relobs}). For the selected
frequencies, we quantify deviations of $\Tatm(z_d)$ and
$\tau(z_d)$ from a simple geometrical scaling $\sim\sec(z_d)$ in
the case of non-planar atmosphere. We also constrain the physical
temperature of the multi-layer atmosphere by introducing a
single-layer equivalent temperature, which we next connect to the
local ground level temperatures by a simple relation
(Sec.~\ref{sec:approximations}). The connection should be readily
useful when constraining atmospheric optical depth due to absorption
and scattering.

Finally, we discuss the sensitivity of a clear sky selection criterion
involving ground-level relative humidity (RH). This criterion can be
used to detect cloudless days from data that only contain measurements
of a few basic atmospheric parameters, such as temperature, pressure and
dew point, which are typically collected by weather balloons and
ground meteorological stations.  By the analysis of archival all-sky
images,
we find that for any given month selecting meteorological data by
the lowest daytime RH can correctly identify
days with a mean
cloud cover below $\sim 0.38$ in 48\% of cases, if one disregards
whether the sky is obscured by low, medium or high level clouds,
and in 76\% of cases if only low and medium level
clouds are considered. We find that reproducing these
true-positive fractions by chance is unlikely at odds greater than $10^4:1$
taking into account the local probability of cloudless skies, and the number of
days used for the analysis (Sec.~\ref{sec:clear_sky_detection}).  We
find that the effectiveness of the estimator is increased during the
hot season (April--September) when the true-positive fraction among
the clear-sky days selected by this algorithm reaches $\sim 89\%$ (for the case when
high level clouds are treated as a clear sky).  We suspect that the
effectiveness of this estimator should be similar in other
locations with a compatible climate.

\section*{Acknowledgements}

BL would like to thank Mike Peel and Marcin Gawro\'nski
for useful discussions on observational pipelines used in
OCRA observations. Also thanks to Scott Paine for comments on
PWV-$\tau$ scaling relations.
Thank you to Peter Wilkinson for comments on the early version
of the manuscript, and to Boud Roukema and to the anonymous referee 
for their useful comments and suggestions.

Use was made of data from the Global Climatology of Atmospheric
Parameters from the Committee on Space Research (COSPAR) International
Reference Atmosphere (CIRA-86) Project.
We acknowledge the use of Integrated Global Radiosonde Archive (IGRA)
data from the Legionowo station.
The data was also sourced from the Canadian Atmospheric
Chemistry Experiment.
We thank Piotr Sobolewski, Brent Holben and Aleksander Pietruczuk for
their effort in establishing and maintaining the Belsk AERONET station.
This research has made use of the AM program (version 7.2) -- a publicly-available
tool for radiative transfer computations at microwave to submillimeter
wavelengths, developed at the Harvard-Smithsonian Center for
Astrophysics.
We also acknowledge use of the 'matplotlib' plotting library \citep{Hunter2007}.

This work was financially supported by the Polish National Science
Centre through grant DEC-2011/03/D/ST9/03373.
A part of this project benefited from the EC RadioNet FP7 Joint
Research Activity ``APRICOT'' (All Purpose Radio Imaging Cameras On
Telescopes).

\bibliography{bibliography} 
\bibliographystyle{mnras}

\bsp	
\label{lastpage}

\end{document}